 \documentclass[final,5p,times,twocolumn]{elsarticle}

\usepackage{amssymb}
\usepackage{booktabs, makecell,tabularx}
\usepackage{algorithm}
\usepackage{algorithmic}
\usepackage{subcaption}
\usepackage{soul}
\usepackage{color}
\usepackage{subcaption}
\usepackage{amsmath,array, graphicx}
\usepackage{tabularx}
\usepackage[table]{xcolor}
\usepackage{comment}
\usepackage{hhline}
\usepackage{todonotes}
\usepackage{multirow}
\usepackage{stfloats}

\begin{document}

\begin{frontmatter}



\title{Explainable Intrusion Detection Systems Using Competitive Learning Techniques}


\author[MSU]{Jesse Ables}
\author[MSU]{Thomas Kirby}
\author[MSU]{Sudip Mittal}
\author[MSU]{Ioana Banicescu}
\author[MSU]{Shahram Rahimi}
\author[MSU]{William Anderson}
\author[ERDC]{Maria Seale}

\affiliation[MSU]{organization={Mississippi State University, Department of Computer Science and Engineering},%Department and Organization
            state={Mississippi},
            country={USA}}

\affiliation[ERDC]{organization={U.S. Army Engineer Research and Development Center},%Department and Organization
            city={Vicksburg},
            state={Mississippi},
            country={USA}}

%\begin{abstract}
%\end{abstract}
\begin{abstract}
    The current state of the art systems in Artificial Intelligence (AI) enabled intrusion detection use a variety of black box methods. These black box methods are generally trained using Error Based Learning (EBL) techniques with a focus on creating accurate models. These models have high performative costs and are not easily explainable. A white box Competitive Learning (CL) based eXplainable Intrusion Detection System (X-IDS) offers a potential solution to these problem. CL models utilize an entirely different learning paradigm than EBL approaches. This different learning process makes the CL family of algorithms innately explainable and less resource intensive. In this paper, we create an X-IDS architecture that is based on DARPA's recommendation for explainable systems. In our architecture we leverage CL algorithms like, Self Organizing Maps (SOM), Growing Self Organizing Maps (GSOM), and Growing Hierarchical Self Organizing Map (GHSOM). The resulting models can be data-mined to create statistical and visual explanations. Our architecture is tested using NSL-KDD and CIC-IDS-2017 benchmark datasets, and produces accuracies that are 1\% - 3\% less than EBL models. However, CL models are much more explainable than EBL models. Additionally, we use a pruning process that is able to significantly reduce the size of these CL based models. By pruning our models, we are able to increase prediction speeds. Lastly, we analyze the statistical and visual explanations generated by our architecture, and we give a strategy that users could use to help navigate the set of explanations. These explanations will help users build trust with an Intrusion Detection System (IDS), and allow users to discover ways to increase the IDS's potency. 
\end{abstract}




\end{frontmatter}

\section{Introduction}
\label{sec:Introduction}

Shifting away from the current trend of black box Intrusion Detection Systems (IDS) can lead to more trustable and credible anomaly detection. Existing methods for AI enabled intrusion detection use Error Based Learning (EBL) algorithms to detect anomalies. EBL refers to models that train through minimizing a \textit{loss} function, generally through the gradient descent algorithm. These models can achieve high detection rates, however they suffer from a few problems. First, these techniques can impose high performative cost. Neural networks can require both high amounts of time and memory to train \cite{neupane2022explainable,iannucci2021performance}. Second, many of these methods have high false positive rates which can harm the overall performance of a real-world IDS \cite{buczak2015survey}. Lastly, these models are not easy to understand and are not innately explainable. Users who use these opaque models do not know how or why a prediction was computed. This can cause a lack of trust and prevent the adoption of AI IDS solutions \cite{Marshan,ables2022creating}.

%As more robust cyber attacks develop across the web, there is a need for more automated defense solutions. The oldest method to protect a network involves a slow and laborious process of discovering an attack and creating its signature. This strategy is known as \textit{signature-based} intrusion detection, where unknown data is compared to discovered signatures. However, a more effective method has begun to gain traction in the field of Intrusion Detection Systems (IDS). \textit{Anomaly-based} detection methods feature an Artificial Intelligence (AI) that learns discrepancies between benign and malicious data packets. The driving force behind this detection method is the use of black box AI algorithms. Black box AI are able to recognize new attacks faster than humans can. This ideally minimizes the amount of damage an attack can cause. Unfortunately, black box AI feature a few problems with regard to intrusion detection. First, many models are computationally intensive and can result in the need for a lot of training time or expensive equipment \cite{neupane2022explainable,iannucci2021performance}. Second, these AI suffer from high false positive rates that can cause harm to external users using a commercial platform \cite{buczak2015survey}. Lastly, users who manage these opaque models do not know how or why a prediction was created which can cause a lack of trust and prevent the adoption of AI IDS solutions \cite{Marshan,ables2022creating}.

eXplainable Intrusion Detection Systems (X-IDS) are a potential solution to the above mentioned problems \cite{neupane2022explainable}. To begin, the Defence Advanced Research Projects Agency (DARPA) defines an explainable system as an AI that can explain the \textit{reasoning} for its decisions, characterize its \textit{strengths and weaknesses}, and convey a sense of its \textit{future behavior} \cite{gunning2019darpa}. There are many methods that can allow current EBL AI models to achieve these tenets. Solutions such as Local Interpretable Model-agnostic Explanations (LIME) \cite{ribeiro2016should}, SHapely Additive exPlantions (SHAP) \cite{lundberg2017unified}, and Layer-wise Relevance Propagation (LRP) \cite{binder2016layer} have the ability to convert black box models into semi-transparent, explainable models. However, the use of these types of solutions comes with downsides and a large performance overhead. Creating surrogate models can be a time consuming and resource intensive process. On the otherhand, white box Competitive Learning (CL) algorithms do not need a surrogate model to be trained since they are innately explainable. Being white boxes, these already transparent algorithms are easy to understand and allow for customizable explanations to be generated. White box CL algorithms are able to meet all of the criteria DARPA has set for explainable systems, and are a good choice for an X-IDS.

%\textbf{ADD EXAMPLES OF CL AND EBL ALGOS}
Competitive learning algorithms differ from the more popular EBL algorithms in a number of ways. Where EBL algorithms, such as deep neural networks and recurrent neural networks, learn by adjusting weights to minimize loss, CL algorithms learn through a competitive process. These CL based techniques, for instance, pits nodes that mimic samples of data against one another. When a node wins in this competition, its weights are adjusted to be similar to the training sample. This enables CL-based techniques to learn by creating abstract representations of data. For IDS datasets, an X-IDS system built using CL models can learn different kinds of attacks and benign behaviors. An important benefit to this type of learning is its ability to be data-mined for visual and statistical explanations. This is a benefit that EBL algorithms lack. See Section \ref{sec:Architectures_Post-Modeling} for more information about CL explanations.

%\hl{An X-IDS system built using CL techniques can...
%} \textbf{use sentenses like this, your goal is to create an X-IDS dont let the reader forget that}

The most ubiquitous CL algorithm one will find is the Self Organizing Map (SOM) \cite{kohonen1982self} and its variants. These algorithms consist of a grid of orthogonally connected nodes that each contain a representation of data. As mentioned previously, these nodes compete against one another to mimic abstract patterns in data. The original SOM algorithm consists of a static map of nodes that confine the training area. The Growing Self Organizing Map (GSOM) \cite{fritzke1995growing} and Growing Hierarchical Self Organizing Map (GHSOM) \cite{dittenbach2000growing} improve upon the original SOM design by growing the map horizontally and vertically. These expansion strategies allow the improved algorithms to learn various abstract representations of data than the original SOM. These algorithms are detailed further in Section \ref{sec:Background}.

There are many benefits to using CL algorithms for an X-IDS. Users of an IDS can use CL models to formulate better responses for \textit{tasks} they must perform. Security analysts can use the model's explanations to better understand attacks in order to better protect their network. Machine learning engineers may also be able to discover deficiencies in the model's logic. Using this knowledge, they can modify the architecture or introduce new training samples to increase its overall effectiveness. Explanations will also lead to increasing the IDS's trust and credibility. This can make users more confident that they will be able to complete their tasks.

In this paper, we detail our customizable X-IDS architecture that leverages CL algorithms to create explanations and accurate predictions. The architecture consists of four phases: Pre-modeling, Modeling, Post-Modeling Optimizations, and Prediction Explanation. Users are encouraged to modify the architecture when they receive explanations that are not helpful. We then compare the accuracy of the CL models used in our X-IDS architecture to other EBL models and discuss explanations generated by the SOM, GSOM, and GHSOM. We find that our models are 1\% - 3\% less accurate than black box EBL models, however, CL models are far more explainable.

%One of the architecture's features is its \textit{user-in-the-loop} attribute that requires users to modify aspects of each of the phases. If a user receives an unsatisfactory explanation, they may modify anything from the preprocessing in the Pre-Modeling phase to the types of explanations in the Prediction Explanation phase. Upon receiving adequate explanations, the user is able to perform a task that is associated with the X-IDS. This could include a Security Analyst protecting a network from attack or a programmer discovering a method to improve the model.

Major contributions presented in this paper are -
\begin{itemize}
    \item An X-IDS architecture featuring three CL-based algorithms, built using DARPA’s guidelines for an explainable system. Self-Organizing Map, Growing Self Organizing Map, and Growing Hierarchical Self Organizing Map models are used to create explanatory visualizations and accurate predictions. We find that CL models are 1\% - 3\% less accurate than EBL models, but CL algorithms are more explainable.

    \item An analysis of statistical and visual explanations for an effective X-IDS. Our X-IDS architecture generates a collection of explainable visualizations ranging from global significance charts to fine-grained feature explanations. Users can use these explanations to understand how and why the model makes decisions.

    \item A pruning process that can significantly reduce the size of the GHSOM model. The GHSOM algorithm can create more maps than a human can have time to understand. Additionally, larger maps can impose a higher performative cost. The pruning process is able to remove less important branches to improve overall prediction speed while losing little accuracy.

    \item A performative analysis of our architecture using traditional accuracy metrics. We compare CL models to existing EBL models using the NSL-KDD and CIC-IDS-2017 datasets. CL models are  1\% - 3\% less accurate than EBL algorithms. Even though they are less accurate, their explainability makes CL algorithms an important tool for X-IDSs.
    
\end{itemize}

%\textbf{update this para with correct headings}
The rest of the paper is outlined as follows - In Section \ref{sec:Related Works}, we discuss background on IDS, XAI, and X-IDS. Section \ref{sec:Background} describes the CL-based algorithms used in this paper and contrasts the CL and EBL techniques. Section \ref{sec:Architectures}, outlines our CL based X-IDS with its architecture presented in Figure \ref{fig:cl_arch}. Section \ref{sec:Experiment} discusses our experimental results. Finally, the conclusion and future work has been discussed in Section \ref{sec:Conclusion}.

\section{Background on Intrusion Detection and Explainability}
\label{sec:Related Works}

Intrusion detection has long been an area where black box models are applied to detect intrusions/anomalies. These black box models tend to boast high detection rates, however, a major issue is their poor explainability. Therefore, the area of intrusion detection can heavily benefit from explanations. In the following subsections, we provide a brief background on Intrusion Detection Systems (IDS), Explainable Artificial Intelligence Systems (XAI), and Explainable Intrusion Detection Systems (X-IDS). 

\subsection{Intrusion Detection Systems (IDS)}
\label{sec:Related Works_Intrusion Detection Systems}

An \textit{intrusion} refers to an action that obtains unauthorized access to a network or system \cite{denning1987intrusion}. %Intrusions can be characterized by a violation of Confidentiality, Integrity, or Availability (CIA). 
An Intrusion Detection System (IDS) consists of tools, methods, and resources that help a Cyber Security Operation Center (CSoC) protect an organization by detecting an intrusion \cite{bace2001intrusion,mcdole2021deep}.
%In this section, we present some related work on Intrusion Detection Systems (IDS), Explainable Artificial Intelligence (XAI), and Explainable Intrusion Detection (X-IDS).
%\subsection{Intrusion Detection (IDS)}
%IDS can be classified as either a host-based IDS or network-based IDS. Host-based IDS are placed on a host system and monitor host activity, incoming and outgoing network traffic \cite{LetouHIDS}. Network-based IDS are built to survey and protect a network of hosts from intrusion \cite{Mukherjee1994NetworkID}. 
IDS can be categorized into operation-based classes, such as signature, anomaly, and hybrid. Signature-based IDS operate by preventing known attacks from accessing a network. The IDS compares incoming network traffic to a database of known attack signatures. Notably, this method has difficulty in preventing \textit{zero-day} attacks \cite{sharma2014evolution}. Anomaly-based IDSs look for patterns in incoming traffic to recognize potential threats and leverage complex AI models \cite{Chandola2009AnomalyDA,mcdole2020analyzing,neupane2022explainable}. A significant drawback of this approach is the tendency for such systems to categorize legitimate, unseen behavior as anomalous. Hybrid-based IDS incorporates the design philosophy of both signature-based and anomaly-based IDS to improve the detection rate while minimizing false positives \cite{szczepanski2020achieving, pang2021explainable}.

Current work on AI enabled anomaly-based IDS can be further divided into black box and white box models \cite{neupane2022explainable}. White box models are considered \textit{easy to understand} by an expert. This allows the expert to analyze the decision process and understand how the model renders its decision. This (semi-) transparent property allows white box models to be deployed in decision sensitive domains, where auditing the decision process is a requirement. White box models may use regression-based approaches \cite{subba2015intrusion}, decision trees \cite{mahbooba2021explainable}, and Self Organizing Maps (SOMs) \cite{langin2011annabell}. Black box models, on the other hand, have an opaque decision process. This opaqueness property makes establishing the relationship between inputs and the decision difficult, if not outright impossible. Black box models comprise nearly all the AI enabled state-of-the-art approaches for IDS, as the focus is traditionally on model performance, not explainability. Examples of popular black box model techniques are Isolation Forest \cite{Liu2008IsolationF}, One-Class SVM \cite{Schlkopf1999SupportVM}, and Neural Networks \cite{ZhangNN}.

As previously stated, state-of-the-art approaches for IDS, as well as machine learning as a whole, focus on model performance through the lens of model accuracy. This focus on model accuracy has driven the development further away from modeling approaches that are transparent or have methods of explainability. In turn, this creates a separation between model inference and the \textit{understanding} of model inference, which gives the inability to confirm model fairness, privacy, reliability, causality, and ultimately trust.

\subsection{Explainable Artificial Intelligence Systems (XAI)}
\label{sec:Related Works_Explainable Artificial Intelligence}

The notion of an Explainable Artificial Intelligence system (XAI) dates back to the 1970s. Moore et al. \cite{moore1988explanation} surveyed works from the 1970s to the 1980s, detailing early methods of explanations. Some early explanations consisted of {canned text} and code translations, such as the 1974 explainer MYCIN \cite{shortliffe1974mycin}. We can find a more current definition of XAI by the Defense Advanced Research Projects Agency (DARPA) \cite{gunning2019darpa}. DARPA defines XAI as `systems that are able to explain their reasoning to a human user, characterize their strengths and weaknesses, and convey a sense of their future behavior'. An XAI system that follows this definition offers some form of justification for its action, leading to more trust and understanding of the system. The explanations from an XAI system help the user not only in using and maintaining the AI model but also helping users complete tasks in parallel with the AI system. Tasks can include doctors making medical decisions \cite{shortliffe1974mycin, holzinger2017we, lindsay2020explainable}, credit score decisions  \cite{chun2021study}, detecting counterfeit banknotes \cite{han2019joint}, advance maintenance \cite{neupane2022temporal}, or CSoC operators defending a network \cite{gunning2019darpa, darpa2016broad,ables2022creating}. 

The current literature consists of many different black box models being used alongside explanation techniques. Common explainer modules for black box models are Local Interpretable Model-agnostic Explanations (LIME) \cite{ribeiro2016should}, SHapely Additive exPlantions (SHAP) \cite{lundberg2017unified}, and Layer-wise Relevance Propagation (LRP) \cite{binder2016layer}. Modern techniques for explaining black box models consist of creating surrogate models that generate explanations either locally or globally. Other methods involve propagating predictions backward in a neural network or decomposing a gradient. More novel approaches have also experimented with making datasets explainable \cite{islam2019domain} or making graphical user interfaces for explainable systems \cite{wu2020feature}.

\subsection{Explainable Intrusion Detection Systems (X-IDS)}
\label{sec:Related Works_Explainable Intrusion Detection Systems}

Explainable Intrusion Detection Systems are still an emerging sub-genre. {The need for explainability in IDS is becoming increasingly necessary. In decision sensitive domains, black boxes obfuscate the decision making process causing a lack of trust in predictions.} The users need to be confident in the predictions or recommendations computed by an IDS. {Understandable explanations allow users to perform their tasks correctly. The stakeholders of an IDS (e.g. CSoC operators, developers, and investors)} are individuals who will be dependent on the performance of the system \cite{neupane2022explainable}. CSoC operators will be performing defense actions based on prediction and explanation results. Developers can use explanations to fortify the model in areas where it is weak. Investors may need explanations to help them make their company's budgeting decisions.

%Explainable systems are becoming more important in decision sensitive domains, as well as to supplement and empower existing knowledge techniques (e.g. data mining, rule-based development) that black boxes obfuscate.

There are many examples of X-IDS being used in research today. A survey by Neupane et al. \cite{neupane2022explainable} describes in detail different X-IDS systems. Many black box implementations have been shown using libraries such as SHAP, LIME, or LRP \cite{wang2020explainable, khan2021new, amarasinghe2018toward}. There have also been more original explanation frameworks, such as one that involves using the CIA triad to generate explanations \cite{islam2019domain}. On the other hand, white box models have also been used to create strong X-IDS architectures. Notable entries have created explainable decision trees and linear regression models \cite{mahbooba2021explainable,subba2015intrusion}. In our previous work \cite{ables2022creating}, we created a proof-of-concept X-IDS architecture that uses a Self Organizing Map (SOM). The architecture, based on DARPA's recommendation \cite{gunning2019darpa}, is meant to be a good starting point for developing explainable IDS systems. Additionally, we demonstrate that the white box SOM algorithm is both more explainable and is comparably accurate to black box models. Both black box and white box methods have proved effective in detecting attacks. However, we believe that the inherently explainable Competitive Learning (CL) based methods are the way forward for future X-IDS.

\section{Background on Competitive Learning (CL) based Algorithms}
\label{sec:Background}

In this section, we briefly describe the theoretical and practical aspects of Competitive Learning (CL) algorithms. CL covers a range of algorithms wherein parts of the model compete against one another to represent one aspect of a dataset. This is opposed to a different method of training models known as Error Based Learning (EBL). {In the following section, we discuss Competitive Learning (CL) and how it compares to Error Based Learning (EBL). We detail various CL based intrusion detection systems in current works. Lastly, we explain the algorithms that we use in our experimentation.} 

%We also dictate a short history of Explainable Artificial Intelligence (XAI) and its use in Explainable Intrusion Detection (X-IDS).

%Parts of the model fight to become a representation of an idea found in a dataset

%The following section will detail what  competitive learning is and why it is \textcolor{red}{superior} to other ML methods. We then detail the three different CL algorithms used in this work: SOM, GSOM, and GHSOM.

%In this section, we briefly describe the theoretical and the practical aspects of Self Organizing Maps (SOM), Growing Self Organizing Maps (GSOM), and Growing Hierarchical Self Organizing Map (GHSOM) and how they can be utilized to detect intrusions and generate explanations.

\subsection{Error Based Learning vs. Competitive Learning}
%\todo{add + intro}

Neural network training algorithms can be divided into a few categories. One of the most popular categories is Error Based Learning (EBL). The core principle behind EBL is optimization. EBL models are trained through a process known as Empirical Risk Minimization (ERM). Through this process, the Machine Learning (ML) algorithm works to minimize a parameter known as `loss', which is a metric that measures how poorly a model predicted a specific sample. If the model is correct, loss is given a value of 0, otherwise, loss will be a value greater than 0. Common loss functions include Binary Cross Entropy, Mean Absolute Error, and Poisson. To make use of the loss function, ML algorithms employ an ERM technique. Gradient Descent (GD) is one of the most well known techniques for this purpose. GD works by calculating the slope or gradient at a given point of a loss function. Normally, this strategy is applied to convex functions, but ML applications are rarely so orderly. After calculating the gradient, GD then takes a \textit{step} down the slope. A \textit{step} can be done for every training sample or a batch of training samples. This changes the weights and biases of a neural network in an effort to lower the loss. This process repeats until the algorithm has converged as close as possible to 0.

%It then applies a learning rate to determine how far it should step to find the next gradient point.

The next set of algorithms that can be used is Competitive Learning (CL). CL consists of unsupervised algorithms where nodes \textit{compete} with one another over the right to activate for input data. There are also a few variants of these algorithms that use probabilistic methods rather than neural networks. CL algorithms follow three tenets \cite{rumelhart1985feature}: (i) all units are the same at the start except for their randomly selected weights, (ii) the `strength' of each unit is limited, and (iii) units compete to represent a sub-set or `cluster' of the input data. Using these tenets, nodes in a CL algorithm can represent abstract patterns or features in data. Nodes compete by being closer to the input data. Generally, this is calculated through euclidean distance. The randomly selected weights are then adjusted to be closer to the input data. There are a few common algorithms that implement this tactic: SOMs, K-Means Clustering, and Expectation-Maximization (EM) mixture modeling \cite{sammut2011encyclopedia}.

One can already begin to see the difference between CL and EBL. Neurons in EBL algorithms represent an activation function rather than mimicking input data. EBLs train towards the goal of minimizing loss from these activation functions. On the other hand, the nodes of CL algorithms contain a vector that is similar to the input data. Training these nodes allow these algorithms to slowly converge toward the inputted samples. Another major difference between these two learning styles is their supervised/unsupervised nature. Many EBLs require a supervised based learning style such that their loss function can be calculated. However, CL algorithms are able to be trained in an unsupervised manner. Data labels are not needed during the training process. Another advantage for CL based methods is that they tend to be inherently explainable. Since the model works to represent clusters in training data, the model can be data-mined for various visual and statistical explanations. The same cannot be said for many EBL methods. As mentioned in a previous section, frameworks such as SHAP \cite{lundberg2017unified} or LIME \cite{ribeiro2016should} may be required to make EBL neural networks explainable. Lastly, the two algorithm sets predict data differently. EBL algorithms predict data using a loss threshold that causes neuron activation {while CL predicts based on proximity.} 

Generating explanations for EBL and CL algorithms also differ. EBL algorithms are categorically known as black box algorithms. It is difficult to discern what process the algorithm took to create predictions. To remedy this, one can use a surrogate model method such as Local Interpretable Model-agnostic Explanations (LIME) \cite{ribeiro2016should}, SHapely Additive exPlantions (SHAP) \cite{lundberg2017unified}, and Layer-wise Relevance Propagation (LRP) \cite{binder2016layer}. These surrogate models create explanations generally through processes such as perturbation or probabilistic set theory. There are two major problems with using these approaches. First, the use of these algorithms is effectively using a black box to explain a black box. The process for generating the explanations can be difficult to understand, so it may be more difficult for users to trust the explanations. Secondly, surrogate generators can be computationally expensive. Not only does one need to train a model, they must then train a surrogate model afterwards. CL algorithms remedy these problems. Since the algorithms are already white box, they can easily be explained. Users can create their own custom explanations that they can trust. Additionally, the computational complexity is generally limited by the size of the CL algorithm's map of nodes. 

\subsection{Competitive Learning used in Intrusion Detection}

\begin{figure*}[ht!]
    \centering
    \includegraphics[scale=.65]{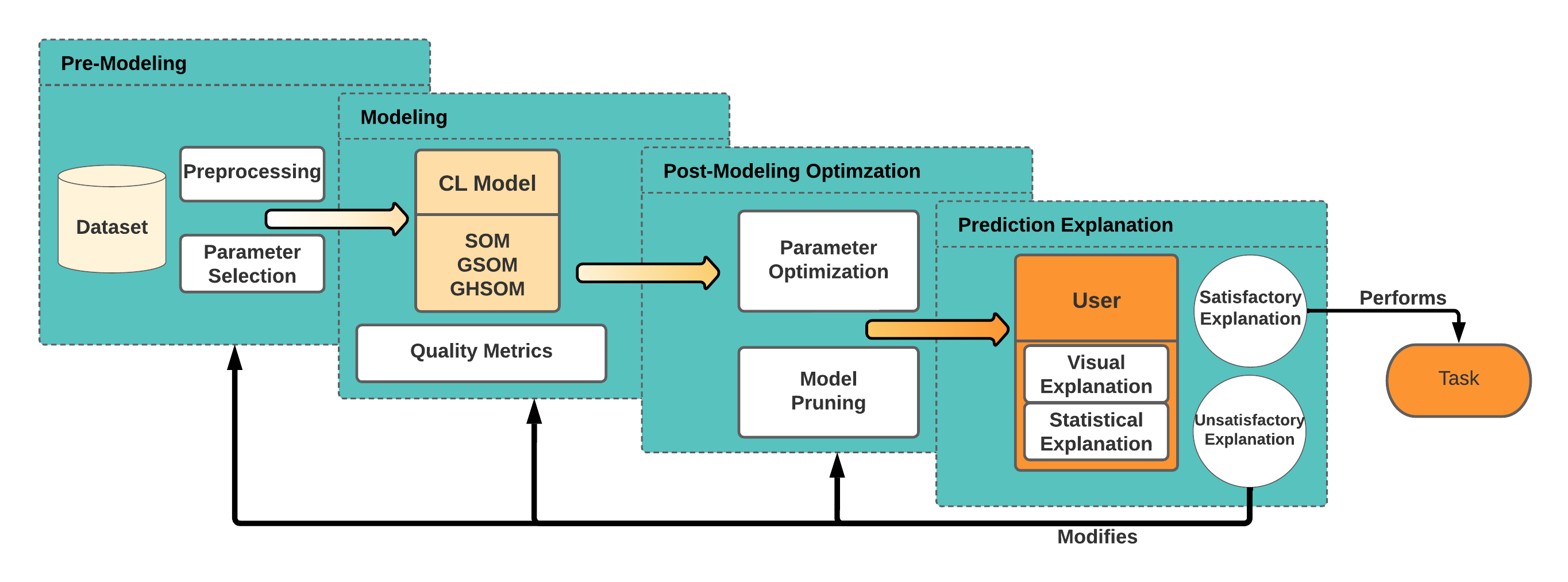}
    \caption{A competitive learning based X-IDS architecture. The architecture is divided into four phases: Pre-Modeling, Modeling, Post-Modeling Optimization, and Prediction Explanation. Each phase contributes to translating raw input data into accurate predictions and useful explanations. Culminating in a user successfully completing an associated task or being required to make changes to previous steps in the architecture.}
    \label{fig:cl_arch}
\end{figure*}

In the past, CL algorithms have been used to create many IDS. These studies focused on building accurate IDS and did not discuss explainability. Among these approaches, SOMs were used to create both host-based \cite{lichodzijewski2002host} and network-based \cite{de2015implementation, pachghare2009intrusion, rhodes2000multiple} IDS. The majority of these methods simply trained a SOM based IDS and illustrated mappings between data points and the associated Best Matching Unit (BMU). The approaches described in \cite{albayrak2005combining,rhodes2000multiple} use multiple SOMs in conjunction with one another to create a more effective IDS. Only one approach \cite{de2015implementation} discussed the false positive rate and accuracy of a SOM-based IDS. Their method for prediction involved assigning a label to BMUs based on the training dataset. Using this approach meant that not all SOM units were assigned a label. The authors utilized Gaussian Mixture Modeling (GMM) to make predictions when a testing sample was similar to an unlabeled unit. In our previous work \cite{ables2022creating}, we created an X-IDS architecture based on DARPA's recommended architecture. One of its main features is having user input for correcting or modifying the model or its explanations. Using this architecture, we were able to achieve an accuracy of 91\% on NSL-KDD and 80\% on CIC-IDS-2017.

In addition, we can look at instances of GSOM-based IDS. A multi-agent GSOM proposed by Palomo et al. \cite{palomo2009self} was created with the goal of being more accurate on datasets with many different attack types. The Growing SOM should be able to continuously grow as it discovers new attack types. Their IDS was able to achieve a 90\% accuracy and a 1\% false positive rate on the KDD CUP 1999 dataset using 38 different attacks. A novel GSOM algorithm was developed in \cite{qu2019statistics} and called Statistics-Enhanced Direct Batch Growth Self-Organizing Map (SE-DBGSOM). One of the goals of using this updated algorithm is to improve the efficiency of inserting new nodes. The authors note that their algorithm improves upon previous GSOMs by reducing the number of `unnecessary' nodes. This improves both runtime and false positive rates. SE-DBGSOM was able to achieve a greater than 99\% accuracy on KDD99 and CICIDS2017 datasets with false positive rates as low as .6\%.

GHSOMs have also made an impact in the field of IDS. One inspiring work that created a GHSOM IDS is from the authors Ippoliti et al. \cite{ippoliti2012ghsom}. They create an Adaptive GHSOM (A-GHSOM) that uses dynamic normalization scaling, an adaptive growth thresholds, and confidence filtering for reducing inconsistent predictions. We can find other works that make other modifications like adding new metrics for numeric and symbolic data \cite{palomo2008new}, enhancing map initialization and weight distribution \cite{salem2013enhanced}, and changing growing conditions \cite{yang2010using}. Many of these implementations were testing using KDD CUP 1999 or NSL-KDD to great effect.

The final two methods for CL algorithms, EM mixture modeling and K-means clustering, have also been used to create effective intrusion detection systems. Both Bahrolo et al. \cite{bahrololum2008anomaly} and Hammad et al. \cite{hammad2022mmm} have created EM mixture model IDS that attempt to categorize the different attacks in IDS datasets. Another work uses a combination of decision trees and EM mixture modeling to create an effective IDS with an accuracy of 94.2\% on the NSL-KDD dataset\cite{bitaab2017hybrid}. There are a few notable works that use K-means clustering or an ensemble featuring K-means to categorize or predict anomalies. Two methods combine K-means with a Naive Bayes classifier to achieve high detection rates on KDD'99 and ISCX 2012 \cite{muda2011intrusion,tahir2016oving}. In Li et al. \cite{li2011anomaly}, their IDS using solely K-means clustering records a detection rate of 82\% on the KDD'99 dataset.

In this work, we focus on the SOM family of CL algorithms. These innately explainable algorithms have been shown in previous works to be highly accurate for intrusion detection. Their \textit{simple-to-understand} nature is conducive to obtaining great explainable algorithms. In addition, the weights generated by the SOM algorithms are easily visualized for explanations. In the following sections, we describe in detail how each of the SOM algorithms operates.

\subsection{Self-Organizing Maps}
\label{sec:Explainable Self-Organizing Maps_Self-Organizing Maps}

Self Organizing Maps (SOMs), sometimes referred to as Kohonen Maps \cite{kohonen1982self,oja1999kohonen}, Kohonen Self Organizing Maps \cite{guthikonda2005kohonen}, or Kohonen Networks \cite{kohonen2007kohonen}, are a class of unsupervised machine learning algorithms. SOMs are comprised of a network of individual nodes, each of which has a feature vector of the same size as the dimension of training data. Some implementations also include a \textit{(x,y)}  coordinate to allow node movement in a two-dimensional (2D) space. This 2D space is typically represented as a square or a hexagonal grid, to easily visualize the represented space. 

 \begin{algorithm}
 
 \caption{POPSOM Algorithm}
  \label{alg:cap}
 \begin{algorithmic}[!h]
 \renewcommand{\algorithmicrequire}{\textbf{Input: }}
 \renewcommand{\algorithmicensure}{\textbf{Output:}}
 \REQUIRE Rows (n), Columns (m), Learning Rate (LR), Total Epochs (T)
 \ENSURE  Weights (W)
 \\
 \textbf{BEGIN}
 \\
  \STATE Allocate n * m element array W
 \\
  \FOR{each node in W}
  \STATE Allocate N element array with random values [0,1]
  \ENDFOR
  \FOR{Each Training Epoch in T}
  \STATE Pick a training sample
  \STATE Find Best Matching Unit using Euclidean Distance
  \STATE Update BMU elements: $w_i=w_i-\lambda * (w_i - i_i)$
  \STATE Update BMU Neighbors
  \STATE Update Learning Rate
  \ENDFOR
 \RETURN $W$ 
 \\
 \textbf{END}
 \end{algorithmic} 
 \end{algorithm}

%Training a SOM model, outlined in Algorithm \ref{alg:cap}, utilizes the following steps: First, a random training sample is picked. 
POPSOM, outlined in Algorithm \ref{alg:cap}, is the SOM algorithm chosen for this work \cite{yuan2018implementation}. It takes four inputs: the number of rows (n), the number of columns (m), the learning rate (LR), and the total number of epochs (T). Radius is also a common parameter that needs to be set in most SOM algorithms, however, POPSOM calculates its initial radius using $n$, $m$, and LR. The $n$ and $m$ determine the size of the map, while $LR$ is how aggressive the model adjusts its weights. The SOM trains for a total of $T$ epochs before finishing. The algorithm begins by selecting a random training sample. Then, the Best Matching Unit (BMU) is calculated by finding the smallest euclidean distance from the training sample to a SOM node. After the BMU is found, it and its neighbors are updated using the formula $w_i = w_i - \lambda * (w_i - i_i)$, where $w$ is the set of BMU weights and $i$ is the set of feature values. $\lambda$ {is the learning rate function that considers the current training iteration, the chosen $LR$, and the distance from the BMU.} Lastly, the learning rate, neighborhood radius, and current iteration numbers are updated. {The function} $\lambda$ {works in a way that it decreases during the course of the training process.}

SOMs have some unique advantages that come with their application. The first is algorithmic simplicity. As shown in Algorithm \ref{alg:cap}, the brevity of the algorithm helps to maintain the desired properties of algorithmic decomposability and tractability. Additionally, due to its unsupervised nature, SOMs can work on a variety of datasets and applications (e.g. data mining and discovery), not just prediction \cite{Ong1999DataMU}. By design, SOMs convert high-dimensional data into a lower dimensional representation. This representation can be topologically clustered and explained through visualizations \cite{pachghare2009intrusion}. One challenge that comes with the application of SOMs is the selection of the size parameters, as the size does not dynamically adjust and there is no \textit{best size} heuristic \cite{Breard2017}. Finally, another challenge with SOMs is their scalability, both in their time complexity, $O(N^2)$, and space complexity. More methods, such as those in \cite{liu2018scalable}, are needed to address these challenges. 

\subsection{Growing Self-Organizing Maps}

SOMs were further improved by dynamically growing the 2D represented space. The Growing Self-Orginaizing Maps (GSOM) was created by Bernd Fritzke in his impactful work \cite{fritzke1995growing}. Their work kept the square architecture common to SOMs, but allowed it to grow by adding columns or rows dynamically. Future implementations would implement systems that allowed the SOM to grow node-by-node rather than with full rows or columns \cite{Alahakoon1998}. The training process of the GSOM is very similar to that of the SOM other than the growing process.

 \begin{algorithm}
 
 \caption{DBGSOM Algorithm}
  \label{alg:GSOM}
 \begin{algorithmic}[1]
 \renewcommand{\algorithmicrequire}{\textbf{Input: }}
 \renewcommand{\algorithmicensure}{\textbf{Output:}}
 \REQUIRE Data Dimension (D), Spread Factor (SF), Learning Rate (LR), Total Epochs (T)
 \ENSURE  Weights (W)
 \\
  \textbf{BEGIN}
 \\
  \textbf{Initialization}
  \STATE Initialize 4 starter nodes with random Weights W [0,1]
  \STATE Calculate Growth Threshold (GT): $GT = -D * ln(SF)$
 \\
  \textbf{Growing Phase}
  \FOR{Each Training Epoch in T}
  \STATE Reset Cumulative Error (CE) for all nodes to 0
  \STATE Present training samples
  \STATE Determine BMU using Euclidean Distance
  \STATE Update BMU and Neighboring weights
  \STATE Calculate CE for all BMUs

  \FOR{all non-boundary nodes}
  \STATE Distribute CE to neighbors
  \ENDFOR
  \FOR{all boundary nodes CE $>$ GT}
    \STATE Grow depending on number of available neighbor positions
  \ENDFOR
  \ENDFOR
 \\
 \RETURN $W$ 
 \\
 \textbf{END}
 \\
 \end{algorithmic} 

 \end{algorithm}

The GSOM algorithm chosen for this paper is the Direct Batch Growing Self-Organizing Map (DBGSOM) \cite{vasighi2017directed}. Its psuedocode can be found in Alg. \ref{alg:GSOM}. It takes three inputs: the dataset's dimensions (D), Spread Factor (SF), and Learning Rate (LR). $D$ is the number of features a dataset has. \textit{SF} determines how quickly new nodes are generated. $LR$ is the same as in the SOM. Another important variable that is not selected by the user is the Cumulative Error (CE). Each node in the GSOM has a $CE$ value. $CE$ is the sum of all the differences between a sample and its BMU. This value slowly accumulates over the course of training.

DBGSOM follows similar tenets as the original GSOM algorithm. The main difference is that it generates new neighboring nodes in a batch process. It is initialized with four starter nodes with randomized weights between 0 and 1. A growth threshold is calculated based on \textit{SF} which is static throughout the training process. After the DBGSOM is initialized, it enters the \textit{Growing Phase}. All nodes have their \textit{CE} reset to 0. Training the GSOM is now similar to training a SOM. Each training sample is presented to the map, and its respective BMU is found. The BMU has its weights and \textit{CE} updated based on the training sample. Additionally, all neighbors of the BMU have their weights updated. After all of the training data has been used to update weights, we find all non-boundary nodes. For each of these nodes, we distribute their \textit{CE} to their neighbors. Lastly, all boundary nodes for which $CE_i > GT$ have a new neighbor node generated next to it.

A major advantage of using this algorithm is the undefined size of the map. SOMs are limited in the fact that they use an unchanging number of nodes. If a map is too small, then different labels from the dataset can begin to merge or take over one another. On the other hand, a larger map may lead to many useless nodes taking up processing time. GSOMs solve this issue by adding new nodes as needed. When the dataset processes a new idea (or a new attack in the case of an IDS dataset), a new set of nodes can be generated with similar weights.

%This algorithm (See: Algorithm \ref{alg:GSOM}) implements two new phases to the training process: a growing phase and a smoothing phase. During the growing phase, training data is presented similarly to the original SOM algorithm. However, a new variable is incremented whenever a BMU has been found. The \textit{Cumulative Error} (CE) value is the difference between the BMU's weight vector and the input vector. This difference is then added to the BMU's CE value. Each node in the GSOM has its own CE, and each of these CEs will grow individually over the course of the training session when it's node is selected as the BMU. CEs grow until one breaches the Growth Threshold (GT). It is important to note that growth can only occur on nodes that are considered boundary nodes. These are nodes that have less than four neighbors. When a boundary node's CE rises above the GT, a new neighbor node is created. New neighbor nodes are initialised with weights that are the same as or similar to their \textit{parent} node. The growing phase ends with the GSOMs Quantization Error is sufficiently small. Next, the smoothing phase begins. This phases main goal is to slowly smooth out the weight vectors of all the nodes and converge the Quantization Error. No new neighbor nodes are created during this phase, which allows for newer neighbor nodes to be trained to be similar to their cluster. 

\subsection{Growing Hierarchical Self-Organizing Maps}

The SOM field would enter another Renaissance in the form of the Growing Hierarchical Self-Organizing Map (GHSOM). The authors, Dittenbach et al., changed the growing algorithm to not only grow horizontally but also vertically (i.e. hierarchically) \cite{dittenbach2000growing}. Each layer of the GHSOM consists of independent GSOMs. For every \textbf{node} in a GSOM, a \textbf{child GSOM} can be generated. The original GHSOM algorithm uses a \textit{Growing Grid} similar to the authors above \cite{fritzke1995growing}. This paper has chosen to use another implantation known as Directed Batch Growing Hierarchical Self-Organizing Map (DBGHSOM) \cite{vasighi2017directed}. A major problem with GSOMs is that the map could grow to be incredibly large, thus causing performance issues and clustering errors. Using vertical, hierarchical growth, the GHSOM can avoid this problem by creating many smaller GSOMs. 

 \begin{algorithm}
 
 \caption{DBGHSOM Algorithm}
  \label{alg:GHSOM}
 \begin{algorithmic}[1]
 \renewcommand{\algorithmicrequire}{\textbf{Input: }}
 \renewcommand{\algorithmicensure}{\textbf{Output:}}
 \REQUIRE Data Dimension (D), Spread Factor (SF), Learning Rate (LR), Total Epochs (T)
 \ENSURE  Weights (W)
 \\
 \textbf{BEGIN}
 \\
  \textbf{Initialization}
  \STATE Same as Alg. \ref{alg:GSOM}
 \\
  \textbf{Horizontal Growing Phase}
  \STATE Same as Alg \ref{alg:GSOM}

  \textbf{Vertical Growing Phase}
  \STATE Calculate the Sum of all CE (SE)
  \STATE Calculate the Vertical Threshold VT = LR $*$ SE
  
  \FOR{All nodes with CE $>$ VT}
  \STATE Create new child DBGSOM
  \STATE Train new child DBGSOM using Alg. \ref{alg:GSOM}
  \ENDFOR
 \RETURN $W$ 
 \\
 \textbf{END}
 \end{algorithmic} 

 \end{algorithm}

The pseudocode for the hierarchical GSOM used in this paper can be found in Alg. \ref{alg:GHSOM}. DBGHSOM has 3 phases: initialization, horizontal growing, and vertical growing. The parameters, initialization, and horizontal growing phases are the same as DBGSOM. A 2-by-2 set of nodes is created and initialized. After the first horizontal growing phase, a vertical growth threshold (VT) is calculated. The vertical growth threshold is a percentage of the total cumulative error of a map. For any node $GT_i > VT$, we create a new DBGSOM. This child GSOM is trained just like its parent.

GHSOMs share some of the benefits that its predecessor has, like being able to dynamically grow. Additionally, they have the benefit of both graphically and abstractly represent data in a hierarchical structure. This form of growth allows for the GHSOM to learn of new attacks as the training data introduces them. The various roots in a GHSOM can be created to have different representations of what constitutes an attack. However, a notable issue with GHSOMs is their ability to grow into thousands of sub-trees and roots, causing performance issues. This problem can be addressed through a pruning process discussed in the next section.

%The pseudocode for this hierarchical GSOM can be found in Algorithm \ref{alg:GHSOM}. It follows the same process as the GSOM where it calculates a Cumulative Error (CE) for each node until a Growth Threshold (GT) is met. There are four different scenarios that apply when a node's CE accumulates above the GT. First, a node with no neighbor slots will evenly distribute half of its CE to neighboring nodes. Second, a node with three open neighbors will grow dependent upon the CE of other Neighboring nodes (NB). Figure \ref{fig: GHSOM Rule 3p} demonstrates the possible NB and open neighbor positions. The position is chosen based off the NB with the highest CE. P1 is chosen if NB1 has the highest CE, likewise for P2 and P3. Third, a node with two open neighbors will follow similar rules. It will try to put a new node next to the highest NB. An important note is when there growing node is between two nodes like in Figure \ref{fig: GHSOM Rule 2p}C, P1 or P2 is randomly chosen. The new nodes weight is $2BO_w - NB_w$ where NB is the closest neighboring node chosen above. The only exception to this is the scenario in Figure \ref{fig: GHSOM Rule 3p} when P1 is not selected, where the new weight is $[(2BO_w - NB1_w) + NBi_w]/2$ where NBi is the neighbor of the new node (i.e NB2 for P2).

\subsection{Model Optimization and Pruning}
\label{optimize}

\begin{figure*}[h!t]
    \centering % <-- added
\begin{subfigure}{0.8\textwidth}
  \includegraphics[width=\linewidth]{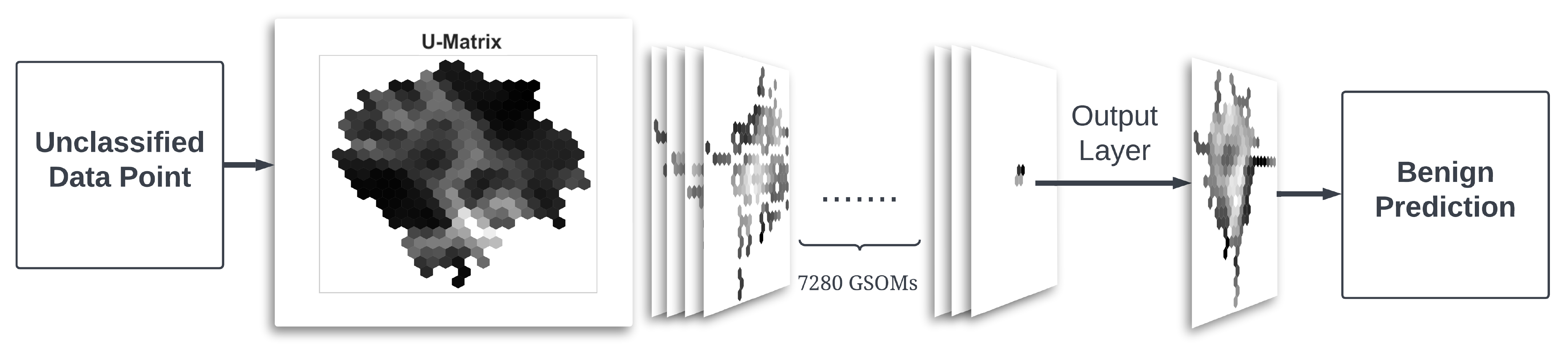}
  \caption{U-Matrices}
  \label{fig:1}
\end{subfigure}\hfil % <-- added
\begin{subfigure}{0.8\textwidth}
  \includegraphics[width=\linewidth]{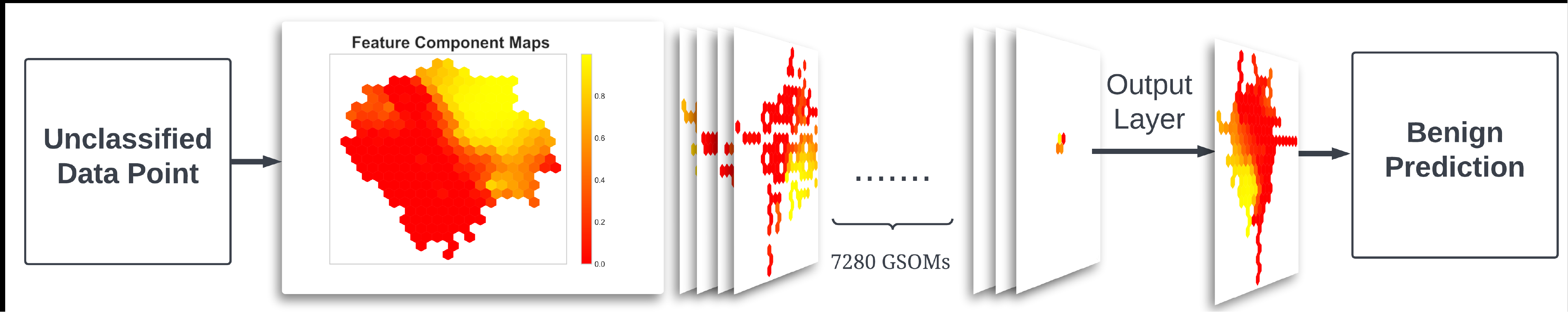}
  \caption{Feature Component Maps}
  \label{fig:2}
\end{subfigure}\hfil % <-- added
\begin{subfigure}{0.8\textwidth}
  \includegraphics[width=\linewidth]{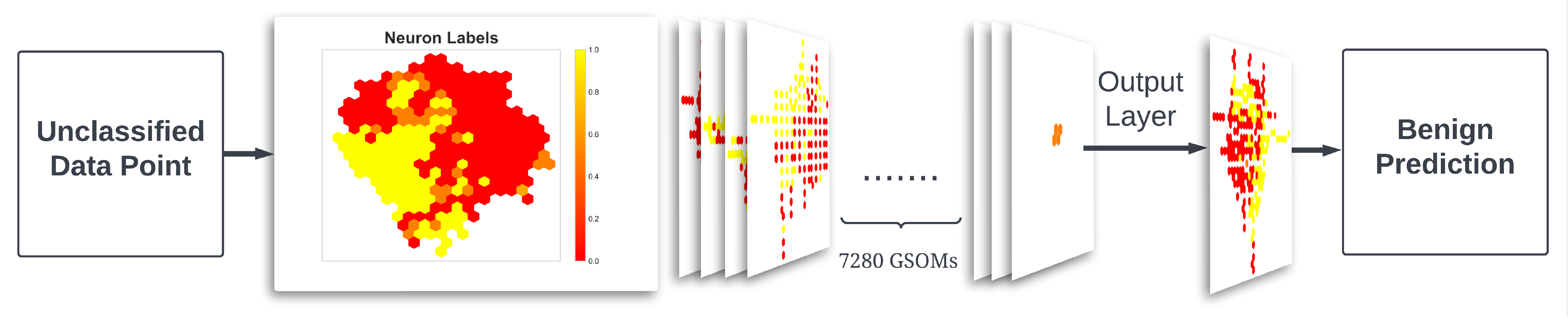}
  \caption{Label Maps}
  \label{fig:3}
\end{subfigure}\hfil

\begin{subfigure}{0.26\textwidth}
  \includegraphics[width=\linewidth]{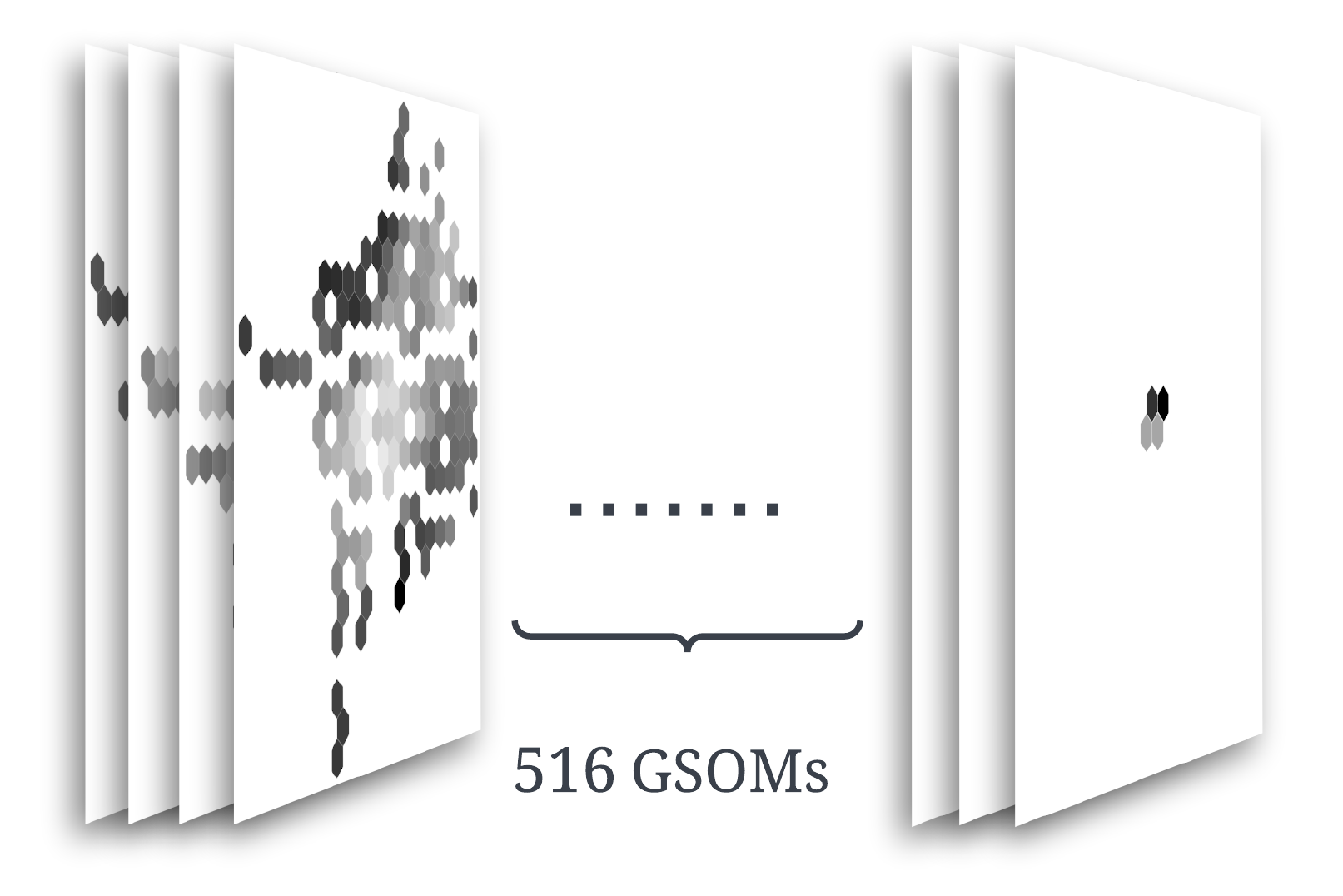}
  \caption{Pruned U-Matrices}
  \label{fig:4}
\end{subfigure}\hfil % <-- added
\medskip
\begin{subfigure}{0.26\textwidth}
  \includegraphics[width=\linewidth]{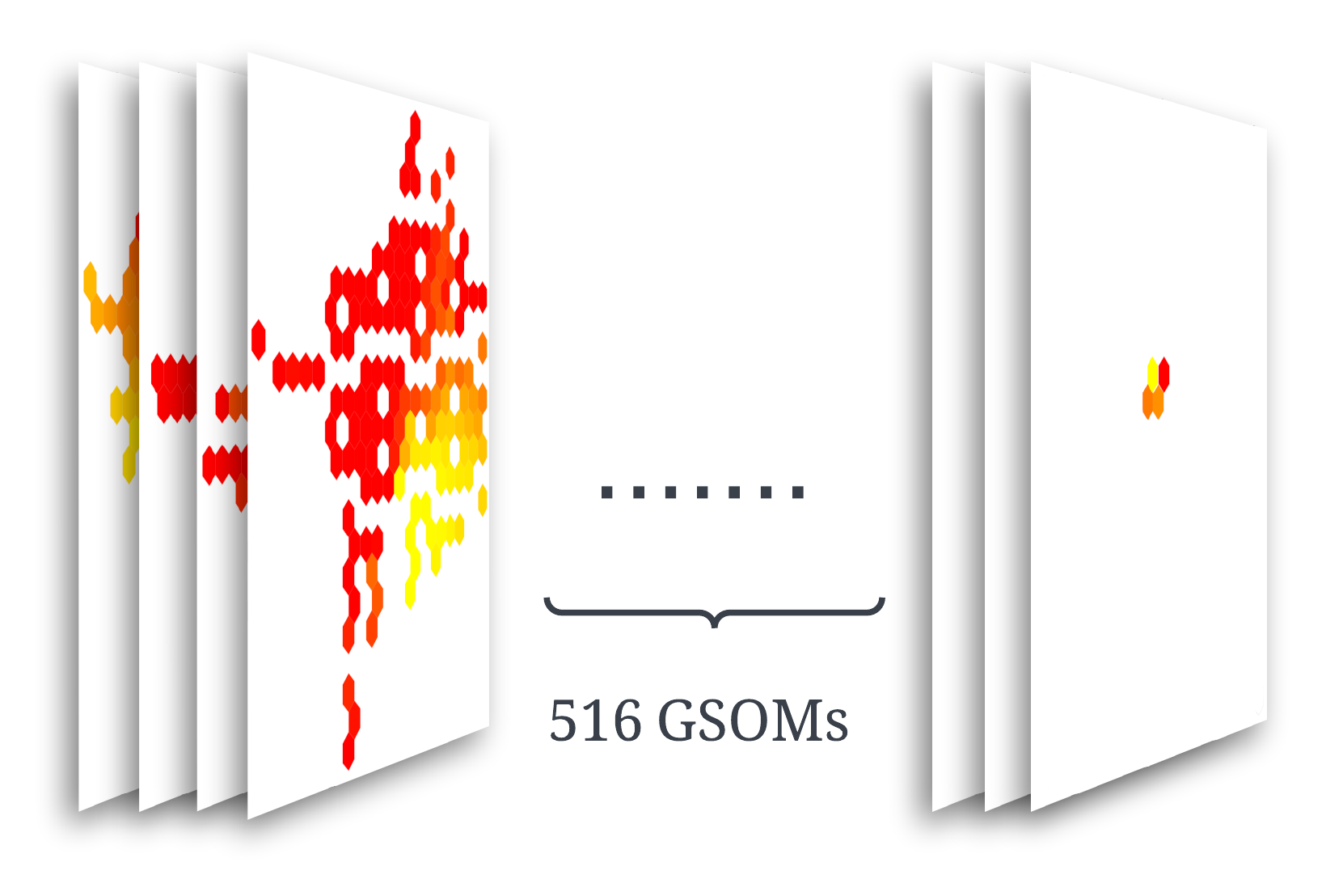}
  \caption{Pruned Feature Component Maps}
  \label{fig:5}
\end{subfigure}\hfil % <-- added
\begin{subfigure}{0.26\textwidth}
  \includegraphics[width=\linewidth]{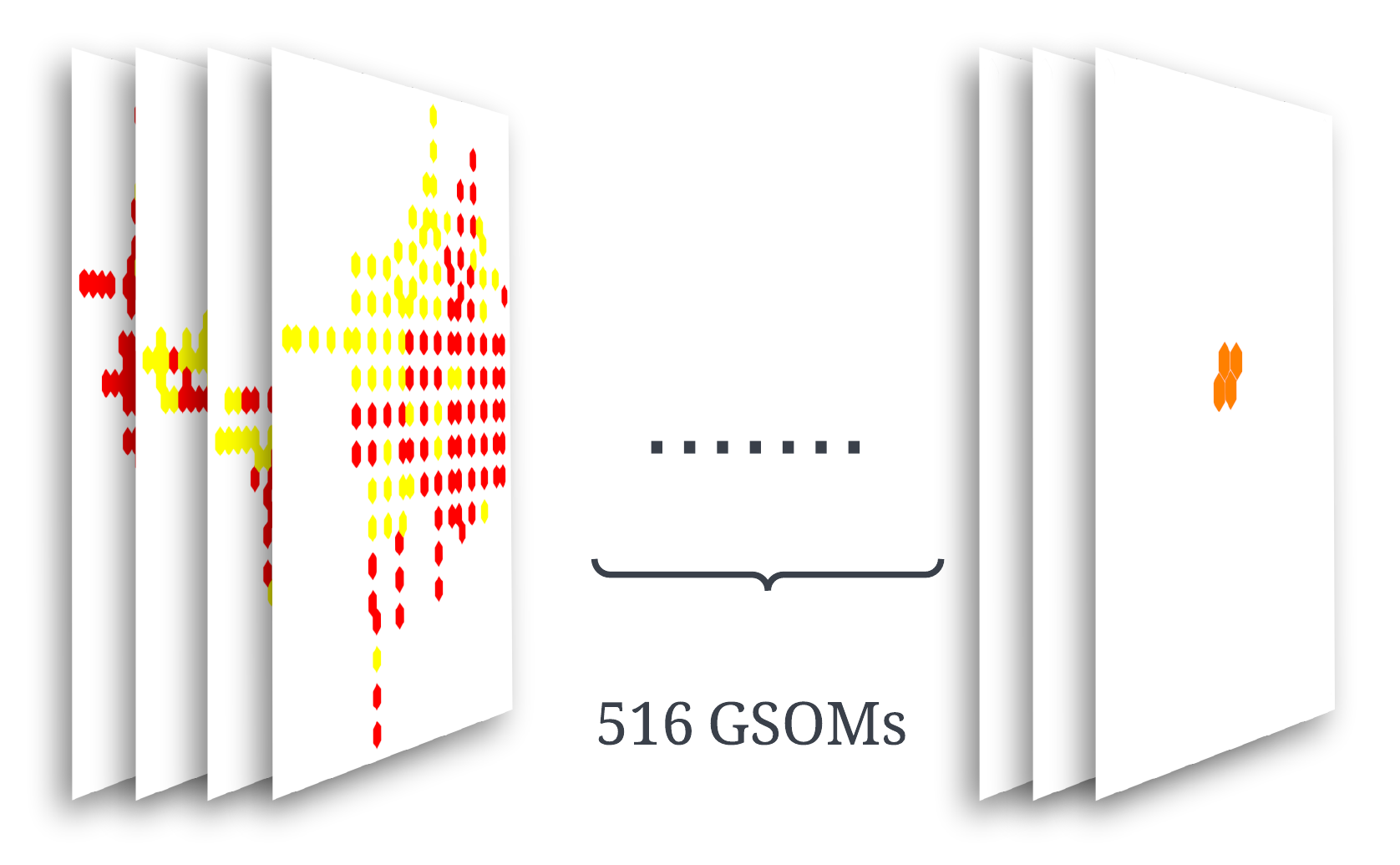}
  \caption{Pruned Label Maps}
  \label{fig:6}
\end{subfigure}
\caption{Visualizations created from GHSOM trained on NSL-KDD. The left hand unslanted visualizations represent the root GSOM. The slanted visualizations represent GSOMs deeper in the hierarchy of GHSOM. Figure (a) shows the Unified Distance Matrices (U-matrices), which shows the distance between nodes with darker areas representing nodes closer together and lighter nodes representing further distances. Figure (b) shows the feature component maps representing the values of specific features on each node in the GSOM. Figure (c) is the Label maps which show the class labels of the node. Figures (d), (e), and (f) represent the pruned versions of the GHSOMs with significantly less network sizes.}
\label{fig: stacked figure}
\end{figure*}
\begin{figure*}[htbp]
    \centering
    \includegraphics[scale=.4]{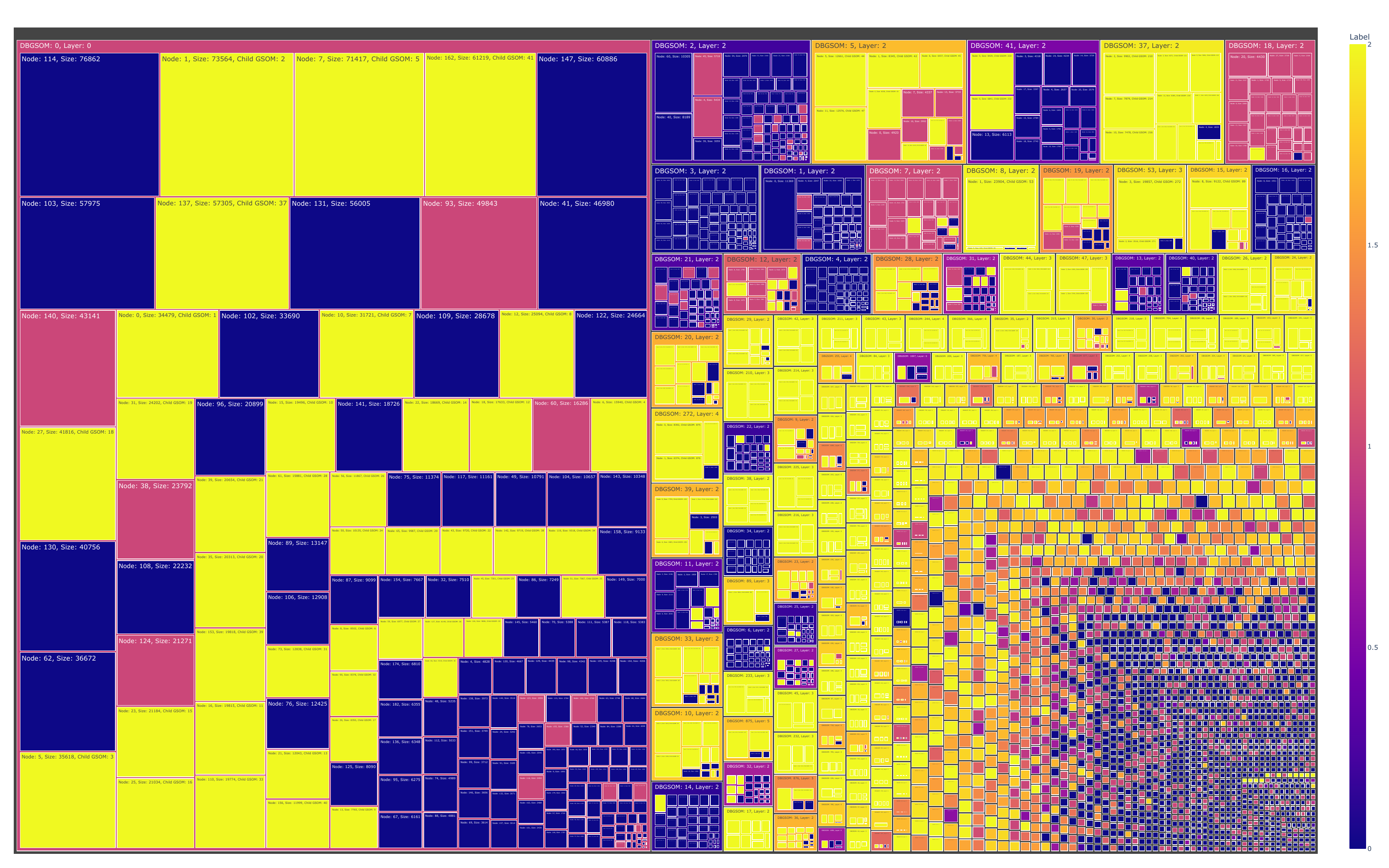}
    \vspace*{-3mm}
    \caption{This figure contains the results from the trained GHSOM on the CIC-IDS-2017 dataset. Since a GHSOM consists of many GSOMs, The tree map diagram displays GSOMs and their nodes. In this tree map, the left half of the map is the root GSOM. Within the root GSOM we can see a mixture of blue, red and yellow nodes. Blue nodes indicate a benign label, red nodes indicate a malicious label and yellow indicate a branch. The size of each node indicates the number of times it was chosen as the BMU.}
    \label{fig: treemap unpruned}
\end{figure*}
\begin{figure*}[htbp]
    \centering
    \includegraphics[scale=.4]{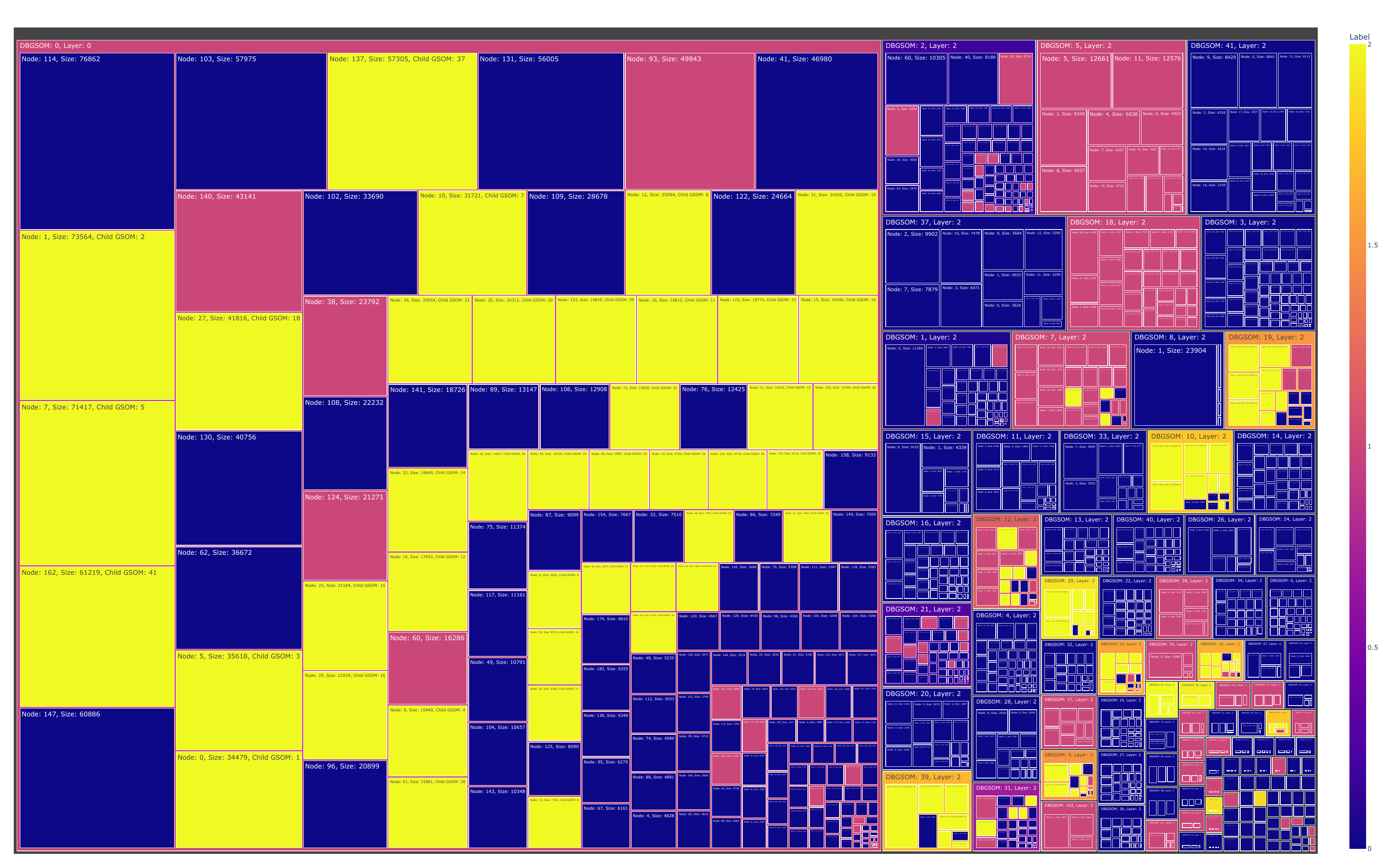}
    \vspace*{-3mm}
    \caption{The tree map generated after pruning the GHSOM from Figure \ref{fig: treemap unpruned}. In the previous tree map, nodes would eventually become too small to see. The pruning process outlined in Section \ref{optimize} allows the user to see the majority of the nodes.}
    \label{fig: treemap pruned}
\end{figure*}

%A tree map containing the results of training a DBGHSOM on the CIC-IDS-2017 dataset. Larger boxes are DBGSOMs that are made up of neurons with each DBGSOM and neuron being labeled with an identifier. Red and blue nodes represent malicious and benign data, respectively. Yellow nodes represent a branch occurring at that node and show the identifier of its child DBGSOM.

One problem with GHSOMs is that it is hard to know when to stop the training process because a branch in a future subtree could provide critical information to the model. So models that grow very large and complex provide good detection rates but at the same time become harder to visualize and explain. A similar issue is found in the decision tree algorithm. Decision trees consist of nodes and branches. Each node contains a feature threshold learned from the training data. The threshold splits the decision making process down separate pathways. When the observed data point reaches a terminal node, a prediction is made using the category of that node \cite{Myles2004}. Researchers sought to develop methods that simplify the decision tree while retaining classification accuracy \cite{Quinlan1987}. A similar method can be used to prune a trained GHSOM.

A pessimistic pruning method based on \cite{mansour1997pessimistic, Kearns1998AFB} can be used on a fully trained GHSOM. From one bottom-up pass through the tree, a decision is made at every node to keep a GSOM or delete it and its children. The pruning decision (see Equation \ref{prune decision}) is based on a comparison of the error rate for the best leaf of a subtree, $e_{bl}$, and the training error rate of the subtree, $e_{st}$, plus a tree complexity penalty, $\alpha$. This method was chosen over other methods as it is computationally efficient, requiring one pass through the tree rather than creating a set of possible subtrees originating from the original tree. All error rates are based on the local input data points mapped to the subtree:

%For a parent node, every child in the parent's subtree needs to have gone through the decision to keep or prune before a decision can be made for a subtree

\begin{comment}
\begin{equation} \label{prune decision}
e_{st} + \alpha \geq e_{bl}
\end{equation}
\end{comment}

\begin{equation} \label{prune decision}
X(e_{st}, \alpha, e_{bl}) = \begin{cases}
e_{st} + \alpha \geq e_{bl} &\text{Remove Branch}\\
\text{else} &\text{Keep Branch}\\
\end{cases}
\end{equation}

\begin{equation} \label{complexity}
\alpha = \sqrt{ \frac{(l+s)\log(n) + \log(\frac{m}{\delta})}{m_v}}
\end{equation}
 
The complexity penalty at any node in the tree can be calculated using Algorithm \ref{complexity}. $l$ is the depth of that node, $s$ is the number of nodes in the subtree, $n$ is the total number of nodes in the tree, $m$ is the total size of the training data, $\delta$ is a confidence parameter between 0 and 1, and $m_l$ is the local size of the data mapped to the subtree. Figure \ref{fig: stacked figure} demonstrates how many possible GSOMs an unclassified prediction may need to search to be labeled, and how the pruning algorithm can help to limit this search. This is beneficial for both performance and explanation generation.

\section{A Competitive Learning based Explainable Intrusion Detection Systems (X-IDS)}
\label{sec:Architectures}
%\hl{The opening para graph can be reduced}
%An X-IDS's main goal is to help stakeholders protect their networks and \textit{understand} various relevant events. The system should act as both a guard and adviser for overall security. When an IDS discovers an intrusion, the user should be notified to prevent a possible attack. Explanations generated by the X-IDS should assist CSoC operators in their mission to protect their organization. 

%\hl{there is no CL in this para or the next}
%\textcolor{red}{deleted first para and made first set of changes to new intro para}

%\hl{You need to change this to say we are using CL based methods to create a proof of concept we are using the SOM family of algorithms. A major rewrite of this intro is needed. }

%in the first stage, in the second stage

Explanations generated by the X-IDS should assist Cyber Security Operation Center (CSoC) operators in their mission to protect their organization. To help achieve this goal, we create the proof of concept Competitive Learning (CL) based X-IDS architecture in Figure \ref{fig:cl_arch}. The proposed architecture is based on DARPA's recommended architecture for XAI systems \cite{gunning2019darpa}. Components of the framework can be changed to suit user's needs. The architecture is abstract enough, such that methods other than CL algorithms can be interchanged to create different X-IDS. The architecture consists of four phases: pre-modeling, modeling, post-modeling optimization, and prediction explanation. In the pre-modeling phase, raw datasets are preprocessed and parameters are selected for the model. In the modeling phase, our CL algorithms are trained and quality metrics are recorded. In our proof of concept system, we are using the SOM family of CL algorithms. In the post-modeling optimization phase, models can then be optimized through various means described below. In the prediction explanation phase, data mining techniques are employed on the resulting models to generate explanatory visualizations that allow users to understand how predictions are generated.

%In the first phase, the model preprocesses raw network data captured into high quality datasets, and selects parameters for the SOM model. Next, the model is trained during the modeling phase. Metrics are recorded to determine the newly trained model's quality. In the optimization phase, a model Lastly, in the post modelling phase, data mining techniques are employed on the resulting map to generate explanatory visualizations that allow users to understand how predictions are generated. In the next subsections, we describe each of these phases in detail.

\subsection{Pre-Modeling Phase}
\label{sec:Architectures_Pre-Modeling}
The pre-modeling phase consists of preprocessing raw datasets and initial parameter selection. The preprocessing for our models includes feature selection and normalization. The feature selection algorithm that we have chosen to use is the `Bayesian probability of significance' \cite{Hamel2012a}, which selects the most relevant features from each dataset. The only other preprocessing that is performed on these datasets is normalization. Feature selection is not used when training the GHSOM. The GHSOM is able to use all of the features in a datset more effectively due to its hierarchical nature. After preprocessing is finished, the new, high-quality dataset can then be passed to the model. The next section details information about the our selected datasets and their usefulness in testing IDS.

%\hl{no transition paragraph, one is need here}

%These datasets are passed through a preprocessing module that normalizes the data. Additionally, the architecture uses Bayesian Probability of Significance \cite{Hamel2012a} to select features. Any feature significance value over a designer selected threshold is included in the preprocessed dataset. The resulting high quality dataset is used during the modelling phase.  

%\subsubsection{Dataset Description}
%\vspace{2mm}
%\noindent \textit{Dataset Description:} 

\subsubsection{Datasets}
In this work, NSL-KDD \cite{Tavallaee2009} and CIC-IDS-2017 \cite{sharafaldin2018toward} are used to test the explainability and effectiveness of our architecture. NSL-KDD is chosen because of its wide use in the literature. %The dataset is a improved version of its 1999 counterpart KDD'99 which was created in the Knowledge Discovery and Data Mining competition. The updated dataset removed many of the duplicate entries which helps reduce biases and improved the testing dataset to be more representative of real-world traffic. 
There are a few major benefits to using the NSL-KDD dataset. First, it allows our method to be compared to other existing methods for IDS. Second, the dataset's relatively small size allows for quick testing and runtime comparisons against larger datasets. On the other hand, CIC-IDS-2017 includes more modern attacks and is useful for testing an unbalanced dataset. %Many datasets, at the time of CIC-IDS-2017's creation, were out of date and not representative of current network data. 
It was synthetically created over the course of  five days to mimic the behavior of 25 users. The use of this dataset, allows us to show that our IDs is compatible with real-world data and to stress-test our systems for various performative metrics. Next, we describe the important characteristics of these two datasets.

\begin{table}[!htb]
    \small
    \setlength{\tabcolsep}{8pt}
\begin{tabularx}{\linewidth}{
       >{\centering\arraybackslash}p{0.15\textwidth}|X
       >{\raggedright\arraybackslash}X  }
        \hhline{|=|=|}
\thead{\textbf{Feature}}  & \thead{\textbf{Description}} \\
        \hline
        Duration & 
        Length of connection \\
        \hline
    %==========================================%
    \makecell{src\_bytes \\ (Source Bytes)} &
    \makecell[l]{Number of bytes from source to\\ destination} \\
    
    \hline
    %==========================================%
    \makecell{dst\_bytes\\(Destination Bytes)} &
    \makecell[l]{Number of bytes from destination\\to source} \\
    
    \hline
    %==========================================%
    \makecell{Count} &
    \makecell[l]{Number of connections to the same\\host as the current connection at a\\given interval} \\
    
    \hline
    %==========================================%
    \makecell{srv\_count\\(Service Count)} &
    \makecell[l]{Number of connections to the same\\service as the current connection at a\\given interval} \\
    
    \hline
    %==========================================%
    \makecell{dst\_host\_count\\(Destination Host\\Count)} &
    \makecell[l]{Number of connections to the same\\destination} \\
    
    \hline
    %==========================================%
    \makecell{dst\_host\_srv\_count\\(Destination Host\\Service Count)} &
    \makecell[l]{Number of connections to the same\\destination that use the same service} \\

        \hhline{|=|=|}
    \end{tabularx}
    \caption{Selected features from the NSL-KDD dataset. These features are used to train the SOM and the GSOM models.}
    \label{table:feature descriptions}
\end{table}

\begin{table}[!htb]
    \small
    \setlength{\tabcolsep}{8pt}
\begin{tabularx}{\linewidth}{
       >{\centering\arraybackslash}p{0.17\textwidth}|X
       >{\raggedright\arraybackslash}X  }
        \hhline{|=|=|}
\thead{\textbf{Feature}}  & \thead{\textbf{Description}} \\
\hline
    %==========================================%
    \makecell{Flow Bytes/s} &
    \makecell[l]{Number of flow bytes per second} \\
    
    \hline
    %==========================================%
    \makecell{Flow Duration} &
    \makecell[l]{Duration of the flow in microsecond} \\
    
    \hline
    %==========================================%
    \makecell{Flow IAT Max\\(Flow Inter-Arrival\\Time Max)} &
    \makecell[l]{Maximum time between two packets\\sent in the flow} \\
    
    \hline
    %==========================================%
    \makecell{Fwd IAT Total\\(Forward Inter-Arrival\\Time Total)} &
    \makecell[l]{Total time between two packets\\sent in the forward direction} \\
    
    \hline

    %==========================================%
    \makecell{Flow Packets/s} &
    \makecell[l]{Number of flow packets per\\second} \\
    
    \hline
    %==========================================%
    \makecell{Destination Port} &
    \makecell[l]{Port the package was destined for} \\

    \hline
    %==========================================%
    \makecell{Bwd IAT Total\\(Backward Inter-Arrival\\Time Total)} &
    \makecell[l]{Total time between two packets sent in\\the backward direction} \\
    
    \hline
    %==========================================%
    \makecell{Fwd Packets/s\\(Forward Packets/s)} &
    \makecell[l]{Number of forward packets per\\second } \\
    
    \hline
    %==========================================%
    \makecell{Flow IAT Min\\(Flow Inter-Arrival\\Time Min)} &
    \makecell[l]{Minimum time between two packets\\sent in the flow } \\
    
    \hline
    %==========================================%
    \makecell{Packet Length Variance} &
    \makecell[l]{Variance length of a packet} \\
    
    \hline
    %==========================================%
    \makecell{Flow IAT Mean\\(Flow Inter-Arrival\\Time Mean)} &
    \makecell[l]{Mean time between two packets\\sent in the flow} \\
    
    \hline
    %==========================================%
    \makecell{Fwd IAT Max\\(Forward Inter-Arrival\\Time Max)} &
    \makecell[l]{Maximum time between two packets\\sent in the forward direction} \\
    
    \hline
    %==========================================%
    \makecell{Idle Max} &
    \makecell[l]{Maximum time a flow was idle\\ before becoming active}\\
    
    \hline
    %==========================================%
    \makecell{Idle Mean} &
    \makecell[l]{Mean time a flow was idle\\ before becoming active} \\
    
    \hline
    %==========================================%
    \makecell{Idle Min} &
    \makecell[l]{Minimum time a flow was idle\\before becoming active} \\
    
    \hline
    %==========================================%
    \makecell{Flow IAT Std Flow\\(Inter-Arrival Time\\Standard Deviation)} &
    \makecell[l]{Standard deviation time between two\\packets sent in the flow} \\
    
    \hline
    %==========================================%
    \makecell{Bwd IAT Max\\(Backward Inter-\\Arrival Time Max)} &
    \makecell[l]{Maximum time between two packets\\sent in the backward direction} \\
    
    \hline

        \hhline{|=|=|}
    \end{tabularx}
    \caption{Selected features from the CIC-IDS-2017 dataset. The features are used to train the SOM and GSOM models.}
    \label{table:feature descriptions cic}
\end{table}

%['duration', 'src_bytes', 'dst_bytes', 'count', 'srv_count', 'dst_host_count', 'dst_host_srv_count', 'label_class']

\noindent\textul{NSL-KDD dataset:} The dataset contains four different kinds of attacks: Denial of Service (DoS), User to Root (U2R), Remote to Local (R2L), and Probing. The DoS is an attack that affects the availability of a service, usually causing that service to be unreachable due to too many connection attempts. However, in general a U2R attack affects the confidentiality or integrity of data. Malicious users, by some means, may gain \textit{root} access to a system and may be able to view or modify data as they please. Similarly, a R2L attack is one where attackers may gain remote access to a machine for which they should not have access. Lastly, probing is a method of information gathering achieved by attackers. This attack looks for known compromised modules that are connected to the internet. While the features (see Table \ref{table:feature descriptions}) of NSL-KDD mainly pertain to tcp/ip packet information, there are also features relating to traffic and content. Traffic features contain information about the duration and amount of connections. Content features contain information about data that the attackers sent in the packet.

%These features are needed to discover R2L and U2R attacks, as these attacks need some form of embedded data to attack vulnerable systems.

%\hl{The feature table has not been referenced in the text.}

\noindent\textul{CIC-IDS-2017:} The dataset includes six types of attacks: Brute Force, Heartbleed, Botnet, Denial of Service, Distributed Denial of Service (DDoS), Web, and Infiltration. A Brute Force attack is a common type of attack whereby a malicious actor tries millions of passwords in an attempt to gain access to a user or administrator account. This type of attack will use resources such as the `rockyou.txt' common passwords list {and can affect all aspects of the CIA triad.} Heartbleed is a {confidentiality} attack that exploits a weakness in the OpenSSL library \cite{carvalho2014heartbleed}. Abusing this bug allows attackers access to encrypted data by reading straight from a compromised system's memory. A Botnet attack refers to the involvement of a set of machines used by an attacker to perform a malicious task. {It can affect confidentiality, availability, and integrity.} The DDoS attacks differ slightly from the DoS attacks in that this type of attack originates from many different hosts. The Web attacks consist of various SQL injections and Cross-Site Scripting attacks {that can affect confidentiality and integrity of data.} Lastly, Infiltration attacks exploit vulnerable software on a user's system to allow an attacker to gain backdoor access {potentially affecting all CIA tenets.} The dataset consists of 80 features generated from network traffic using CICFLowMeter \cite{lashkari2017cicflowmeter}. A quick reference for the selected CIC-IDS-2017 features can be seen in Table \ref{table:feature descriptions cic}.

\subsubsection{Model Parameters}

The SOM parameters consist of: $n$ x $m$ rows and columns, the number of training epochs, and a learning rate (LR). Picking the size of $n$ and $m$ is one of the most important decisions. If the values are too small, clusters will not separate and can begin to merge. If the values are too large, proper clusters may not form at all {and resources can be wasted.} Determining what this parameter should be is done through trial and error. Similarly, the number of training epochs can cause significant performance increases. Both scenarios, the one with too many epochs and the one with too few epochs, can cause overfitting and underfitting, or simply waste time. Lastly, LR values affect the learning rate function. A higher value will cause larger changes when adjusting Best Matching Units (BMUs) towards training samples. If the learning rate is set too high, BMU values will mimic singular instances of data. This leads to a less abstract understanding of the data that can cause decreases in accuracy. Too low a value will cause the model to not learn at all. In our testing, we found that a moderately low LR worked best for our datasets.

GSOM and GHSOM contain different parameters than the SOM. These algorithms no longer need to have the number of rows and columns defined. They always start with a 2 x 2 set of nodes. A new, important parameter is the Spread Factor (SF). The SF determines how much Cumulative Error (CE) (see Section \ref{sec:Background}) is needed to create a new, horizontal or vertical node. Smaller SF values cause more node generation. LR is used for the same purpose as in the SOM. It determines how much a BMU will change with regard to the training samples. However, the GHSOM uses LR to also dictate the vertical growth threshold. The pruned GHSOM includes an additional parameter $\delta$ which determines how aggressively the model is pruned. Table \ref{table: parameters} contains all of our experimental settings.

%All of our experiments' parameter settings can be viewed in Table \ref{table: parameters}.

%To view all of our experiment's parameter settings, please view Table \ref{table: parameters}.

%The NSL-KDD dataset, after being preprocessed, is left with features in Table \ref{table:feature descriptions}. The features consist of \textit{Duration}, \textit{Source Bytes}, \textit{Destination Bytes}, \textit{Count}, \textit{Service Count}, \textit{Destination Host Count}, and \textit{Destination Host Service Count}.

%To make understanding the significance of the explainability results easier, we will include a description of the selected features from each dataset. A more succinct view of these descriptions can be found in Table \ref{table:feature descriptions cic}.

Data preprocessing and parameter selection are difficult processes that require many iterations to attain the best results. In our work, we have determined that minimal preprocessing is needed to attain high performative results. Initial parameter selection can be done by hand, while later, the user relies on search algorithms depending on the architecture. Both GSOM and GHSOM have been optimized using a process outlined in Section \ref{sec: post-modeling optimization}. The next phase will take the newly preprocessed dataset and selected parameters to create well trained models. 

%The process for selecting parameters can either be done by hand or with an algorithm.

% \usepackage{multirow}
\begin{table}[]
\begin{tabular}{l|l|l|l}
%\cline{2-4}
\hhline{|==|=|=|}
                                                     & \textbf{Parameter} & \textbf{NSL-KDD} & \textbf{CIC-IDS-2017} \\ \hline
\multicolumn{1}{c|}{\multirow{4}{*}{\textbf{SOM}}}   & n                  & 18               & 18                    \\
\multicolumn{1}{c|}{}                                & m                  & 18               & 18                    \\
\multicolumn{1}{c|}{}                                & LR                 & .3               & .3                    \\
\multicolumn{1}{c|}{}                                & Epochs         & 1000             & 1000                  \\ \hline
\multicolumn{1}{l|}{\multirow{3}{*}{\textbf{GSOM}}}  & LR                 & .006                & .006                     \\
\multicolumn{1}{l|}{}                                & SF                 & .9                & .9                     \\
\multicolumn{1}{l|}{}                                & Epochs         & 100             & 40                  \\ \hline
\multicolumn{1}{l|}{\multirow{4}{*}{\textbf{GHSOM}}} & LR                 & .006                & .006                     \\
\multicolumn{1}{l|}{}                                & SF                 & .3                & .3                     \\
\multicolumn{1}{l|}{}                                & Epochs         & 100             & 40   
            \\
\multicolumn{1}{l|}{}                                & $\delta$        & .3            & .3  \\ \hhline{|=|=|=|=|}
\end{tabular}
\caption{The selected parameters for each CL model. $n$ and $m$ determine the amount of rows and columns the SOM will have. LR is the learning rate that will affect node weight adjustments. SF is the spread factor that affects how often new nodes form in GSOM and GHSOM. Epochs are the total number of training iterations a CL model will go through. $\delta$ is the confidence interval used to prune the GHSOM.}
\label{table: parameters}
\end{table}

\subsection{Modeling Phase}
\label{sec:Architectures_Modeling}

Using the high quality dataset and the parameters selected in the pre-modeling phase, we can train the set of CL models. We utilize a subclass of CL models described in Section \ref{sec:Background} which includes three variants of the self organizing map algorithm: the Self Organizing Map (SOM), the Growing Self Organizing Map (GSOM), and the Growing Hierarchical Self Organizing Map (GHSOM). These algorithms create clusters mimicking input data, and in doing so, they create a map that can be data-mined for explanatory purposes. The SOM is the most basic of the three algorithms. It consists of a grid of nodes with no logic to grow and change shape. While, this allows the algorithm to be more easily understood, it can lead to poorer {performative results.} The GSOM and GHSOM algorithms address this issue by allowing the map of nodes to grow horizontally and vertically. These algorithms are able to understand more complex structures in data, and their growing nature helps to accommodate these new structures. Please see Section \ref{sec:Background} for a more in depth description of these algorithms.

%Various quality and performative metrics are recorded here to gain insights into how the model performs. For SOM like algorithms, metrics such as topographical and quantization error can be recorded. These metrics can help users know that the model has been trained well. Additionally, metrics such as F1-score, accuracy, precision, recall, false positive rate, and false negative rate can also be noted. All of thees metrics combined should be used to determine if the model is performing acceptably. To see some of these metrics in action, one can view our experimental results in Section \ref{sec: performative results}.

%The modeling phase begins by training the  model using the high quality dataset prepared during the pre modeling phase. For this paper, we use the POPSOM implementation \cite{yuan2018implementation}. Training parameters include total training iterations, learning rate, and SOM size. At the end of the training session, the model will be tested to produce topographical error, quantization error, F1-score, precision, recall, and a confusion matrix. The confusion matrix can be used to determine both the false positive and false negative rate for the model. The quality metrics are used to determine if a model has been sufficiently trained to generate explanations.

%\vspace{2mm}
\subsubsection{Model Evaluation Metrics \& Techniques}
There have been various metrics and measures proposed to evaluate the quality of a trained SOM. These include quantization error, topographic error, embedding accuracy, and convergence index. The quantization error was used by Kohenen \cite{Kohonen1998}, and measures the average distance between nodes and the data points. {The topographic error measures how well preserved features are in the low dimensional output space. It is measured} by evaluating how often the BMU and the second BMU are next to each other \cite{Breard2017,Lampinen1992}. The map embedding accuracy is similar to the quantization error. This metric measures how similar the distribution of the input data is with respect to that of the SOM units \cite{Hamel2016}. In order to measure both topographic preservation and distribution similarity between the input and SOM units, the convergence index was proposed to be a measure that linearly combines the map embedding accuracy and the topographic error \cite{Tatoian2018}. {Performative metrics} are also important to include in an IDS architecture. These metrics include accuracy, F1-score, false positive rate, and false negative rate. We also opt to include training time and prediction speed since they can play an important role in intrusion detection. The experimental results using these performative metrics can be view in Section \ref{sec:Experiment}. These measurements allow the architecture to be compared to the architecture of other existing IDS.

%Performative metrics} for our architecture can be viewed in Section \ref{sec: performative results}. These measurements allow the architecture to be compared to other existing IDS.

%To measure how much the features of the input space have been preserved in low dimensional output space, a topographic error is used.

\subsection{Post-Modeling Optimization Phase}
\label{sec: post-modeling optimization}

Selecting the best parameters for a ML model is an important challenge. {However, d}oing such work by hand is a time consuming process. This is especially true when training times begin to scale higher. Luckily, there are methods that can automate this process. Additionally, some models may also benefit from post-processing. Many tree based models will use a form of post-processing known as pruning to reduce the chance of overfitting and speed up decisions. In this work, we have chosen two techniques for post-modeling optimization. A Bayesian search process is employed to find the parameters for improved performative results, and a pruning technique is used to optimize the size of the GHSOM.

\subsubsection{Parameter Optimization}
There are a few notable methods that can be used for parameter optimization. For this work, we considered  Grid search, Random search, and Bayesian search. The differences in these search algorithms relate to how each selects its next set of test parameters. Grid search acts like a brute force approach. It tests as many combinations of parameters as possible. This process is time consuming, but one is more likely to find the best parameters for their model. Random search differs from this approach by sampling a specified number of time from a range of values. This approach limits the number of options that are tested but does not check the entire space of parameter combinations. The last option and the one chosen for our architecture is the Bayesian search. This method limits the search space by creating a surrogate probability model. It makes informed decisions about each set of parameters tested. On average, Bayesian search is able to find a better set of parameters faster than the other two algorithms. The trade-off is that it may not find the best set of parameters as it doesn't search the entire parameter space.

%This allows for the algorithm to find a better set of parameters faster than the other two search algorithms on average.

\subsubsection{GHSOM Model Pruning}
Model pruning can be a valuable resource for certain algorithms. GHSOMs can benefit from pruning by reducing the number of nodes needed to match against. Another notable reason to prune a GHSOM is for easier visualization. GHSOMs can grow into hundreds or thousands of branches making visualization a daunting task. The goal of pruning a tree is to limit its size while also retaining as much information as possible. As mentioned in Section \ref{sec:Background}, we use a pessimistic pruning approach. It is a bottom-up approach that determines whether each node should be kept. To prune a node, an error rate comparison is made. When the error rate for a node is high in comparison to the average error rate, it is removed from the tree. Since this is a bottom-up approach, a node and all of its children can be removed. The visualization changes can be seen in Figures \ref{fig: treemap unpruned} and \ref{fig: treemap pruned}.

Optimized models are important for both detection rates and performance. The benefits from this phase can be seen in Table \ref{results table}. Using parameter optimization, we are able to increase the accuracy of our models, while the pruning process allows for faster predictions. These are both important factors when an system is being used of network security.

%The optimizations chosen for our models include parameter optimization and model pruning. Parameter optimization is an iterative process where a model is trained with different parameters with the goal of increasing detection rates. This paper chose to use the Bayesian Search module found in Scikit-Learn. The main reason for this choice was to limit the search space. It uses previous results from its search to create a surrogate probability model. Rather than searching the entire space, the search algorithm references the surrogate model to attempt to make informed choices about the next set of hyper parameters. The second method used to optimize the GHSOM was a pruning algorithm mentioned in \ref{optimize}. Pruning is a useful tool for improving the prediction speed and lowering the memory requirements of a GHSOM.

\subsection{Prediction Explanation Phase}
\label{sec:Architectures_Post-Modeling}

Once the modeling and optimization phases have been completed, and the quality metrics have ensured that the model is a good representation of the data, the model can be used to perform a variety of explainability and visualizations. The models themselves are lists of SOM nodes and the weights associated with these nodes. Visualizations include creating local and global explanations, U-Matrices, and feature heatmaps. Users can use explanations to perform tasks to better defend the network. When a user receives a subpar explanation, the user can modify to the architecture where needed to help bolster the X-IDS. 

\subsubsection{Local and Global Explanations}
    
Global and local explainability can be achieved by examining important features of the trained SOM, and utilizing this information to generate an explanation for a specific data instance classification or cluster classification \cite{wickramasinghe2021explainable}. Global significance for NSL-KDD is shown in Figure \ref{fig: nsl-global} with higher values denoting that a feature has a higher probability of being important. The algorithm chosen to determine this variance was `Bayesian probability of significance' \cite{Hamel2012a}. Higher variance features increase the probability that a model will capture the dataset's structure. Through this graph, an analyst can understand which features are important to the overall SOM structure, allowing them to examine predictions at a local level based on globally important features.

Figure \ref{fig: nsl-local} shows the GSOM local explanations for a prediction on the NSL-KDD dataset. Each feature has a value representing the significance. Significance (S) is a calculation involving the \textit{min-maxed} distances from a BMU (See Section \ref{sec:Background}) inverted so that higher values are more important. The formula can be seen in Formula \ref{formula: significance}. In this example, we can see the features with the largest impact on a prediction: destination (dst) host count, duration, and destination (dst) bytes (see Table \ref{table:feature descriptions}). These features were the closest to the BMU, therefore, they played a large role in computing the predicted value. Seeing the specific features that influence predictions provides insight into samples labeled as malicious or benign and can further help users determine the reason for incorrect predictions. These features can also be further investigated with feature value heat maps.

\begin{equation} \label{formula: significance}
    S = 1 - (\frac{X - X_{min}}{X_{max} - X_{min}})
\end{equation}

\subsubsection{Unified Distance Matrix (U-Matrix)} The U-Matrix visualizes the distances between neighboring SOM nodes. With distances shown as a color gradient, nodes far apart will create light boundaries while areas with similar nodes will be darker. This can visually represent the natural clusters of input data. To enhance the standard U-Matrix, the starburst model uses connected component lines of nodes overlaid on the matrix to better represent clusters \cite{Hamel2012}. For a labeled data set, the user is able to visualize each BMU along with the associated label. Figure \ref{fig: nsl-U-Matrix} shows clear clusters with boundaries separating malicious (1) and benign (0) behavior. Using this information a users can investigate more visualizations and feature importance values to gain an understanding of why certain malicious network activities are being grouped together.
\subsubsection{Feature Value Heat Map}
{A heat map applied to a feature shows general trends that a feature has on a model, in this case the entire SOM model. SOM feature values are represented from 0 to 1, and the heat maps denote this with darker and lighter values, respectively. An example feature value heat map can be found in Figure \ref{fig: nsl-Feature}. In this example, the `dst bytes' feature has a cluster of higher values in the {top-right} corner, while the rest of the SOM consists of lower values. Users can use this information to form conclusions about the model. Feature value maps are more powerful when multiple are viewed at a time. The U-Matrix chart can then be referenced to make general decisions about the model. The heat maps work well as a fine-grained global explanation that helps users to understand the overall model.} 

\subsubsection{Users Performing Tasks}

An important component of our architecture is its \textit{user-in-the-loop} system. A `user' is a network's stakeholder. There can be many kinds of stakeholders for an IDS. AI engineers who implement and maintain the X-IDS architecture, security analysts who protect the network, and investors manage security expenses. Tasks are performed with the goal of protecting the network and are enhanced by the X-IDS's generated predictions. 

When a satisfactory prediction has been created, a user can perform their task. Satisfactory explanations will cause the user to be able to perform their tasks more effectively. However, not all explanations will be useful. When an unsatisfactory explanation is created, a user can use that explanation to make changes to the parts of the architecture. This could be accomplished by changing how datasets are preprocessed, choosing a new ML model, modifying optimizations, or creating a new style of explanation. 

Prediction explanation plays a pivotal role in an X-IDS architecture. Visualizations and statistics are critical for creating actionable explanations for a network. Local explanations can help programmers and security analysts to fine-tune the model, while global explanations can be used {by investors} to understand the model at a high level. Users can use predictions to either defend the network, or fortify the intrusion detection system. In the next section, we demonstrate the performative results of our models using the above described architecture.

\section{Evaluation \& Experimental Results}
\label{sec:Experiment}

\begin{figure*}[ht]
    \hspace{15pt}
     \begin{subfigure}{0.2\textwidth}
         \hspace{15pt}
         \subcaptionbox{NSL-KDD Local Anomaly Explanation\label{fig: nsl-local}}{%
         \includegraphics[scale=.6]{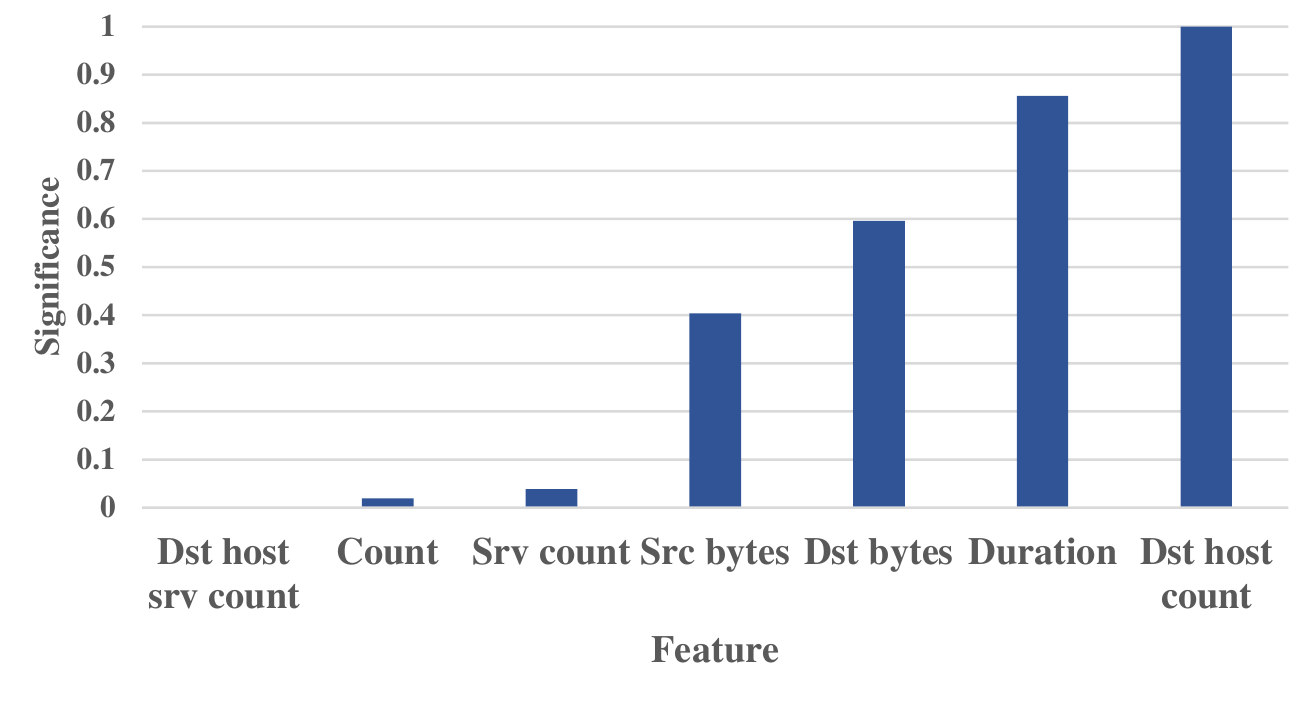}}
         
     \end{subfigure}
     \hspace{130pt}
     \begin{subfigure}{0.3\textwidth}
         \subcaptionbox{NSL-KDD Global Feature Significance\label{fig: nsl-global}}{%
         \includegraphics[scale=.6]{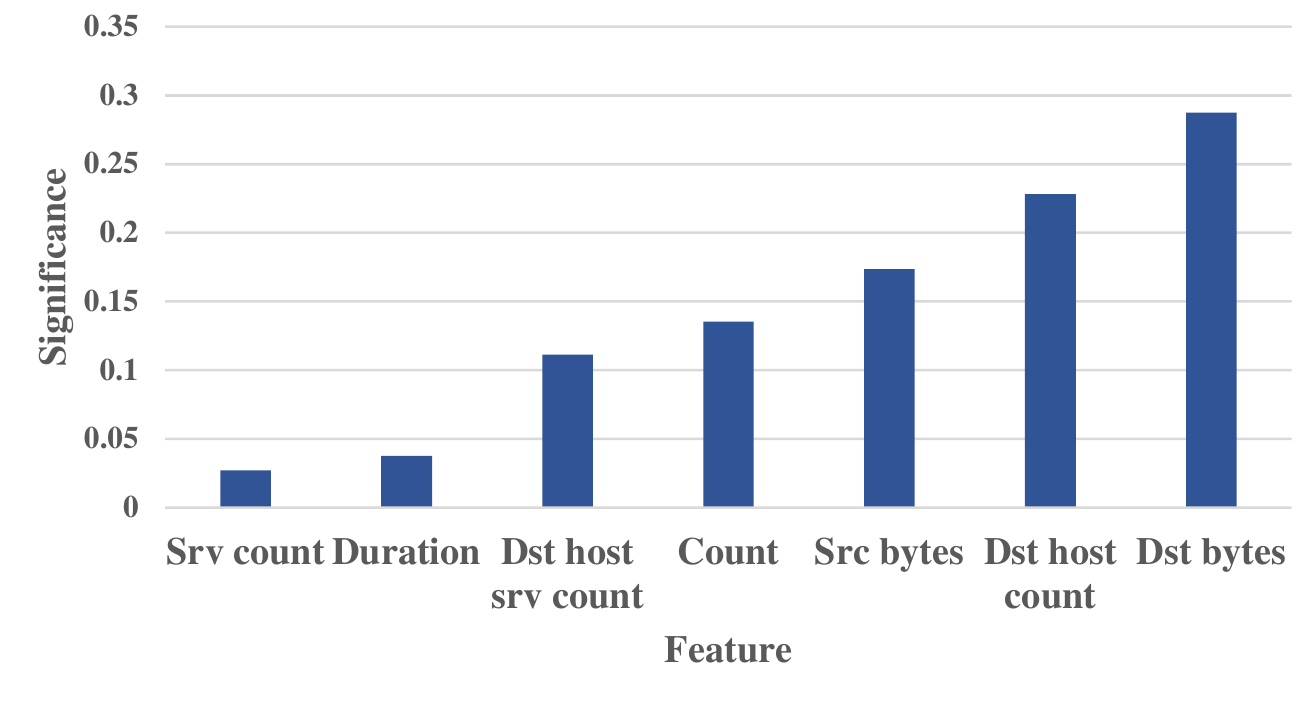}
         }
     \end{subfigure}
     \vfill
     \hspace{10pt}
     \begin{subfigure}{0.3\textwidth}
     \subcaptionbox{\mbox{CIC-IDS-2017 Local Anomalous Explanation}\label{fig: cic-local}}{%
         \includegraphics[scale=.6]{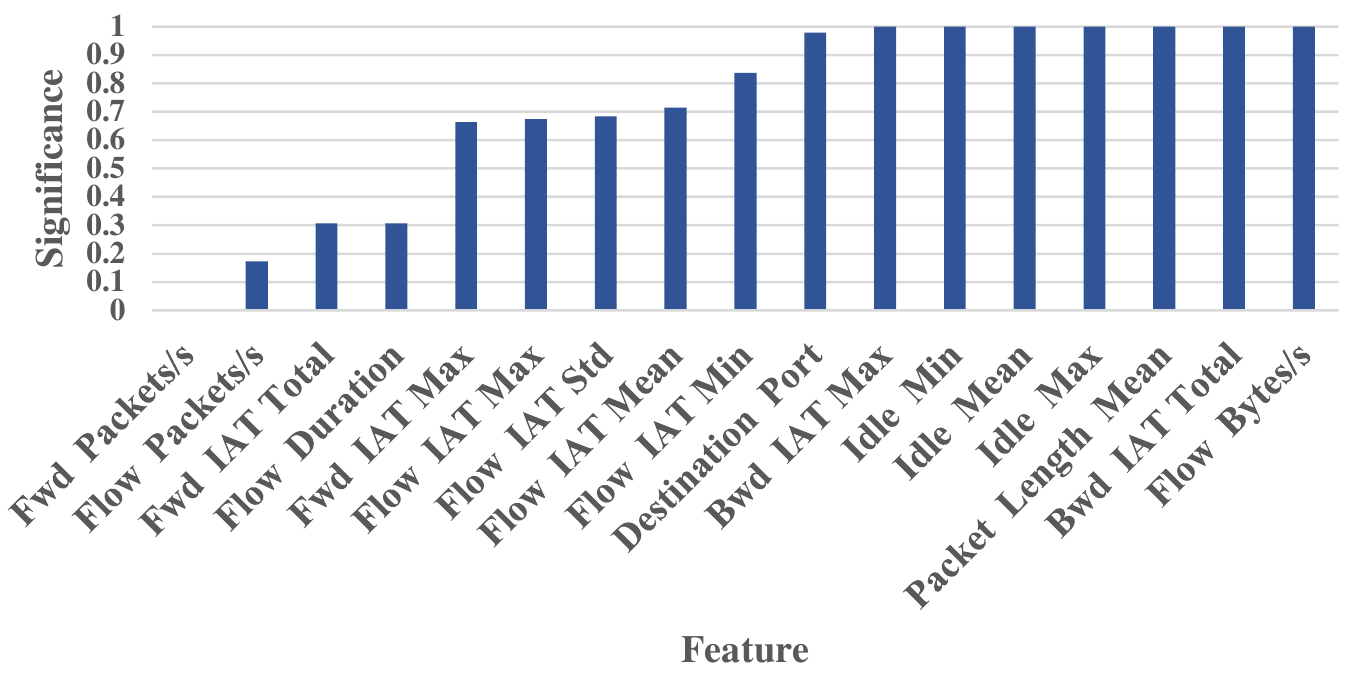}
         }
         
     \end{subfigure}
     \hspace{80pt}
     \begin{subfigure}{0.3\textwidth}
         \subcaptionbox{\mbox{CIC-IDS-2017 Global Feature Significance}\label{fig: cic-global}}{%
         \includegraphics[scale=.6]{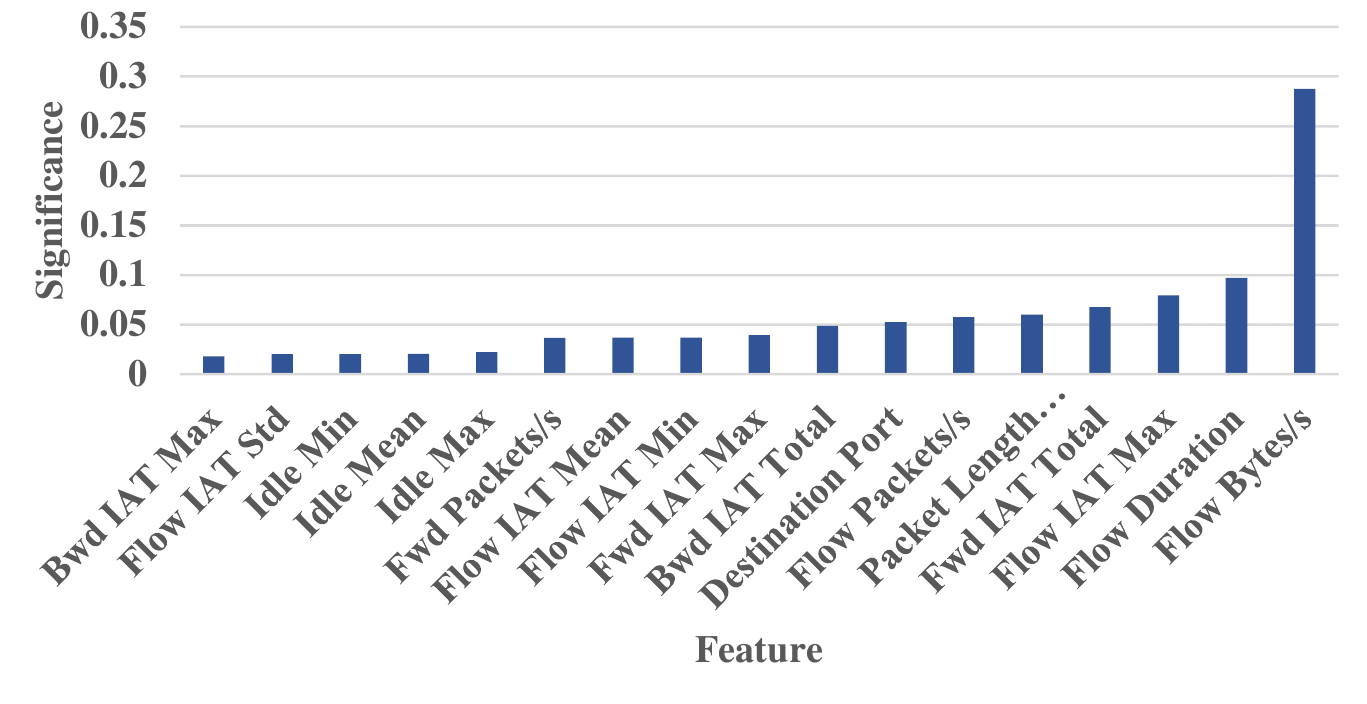}
         }
     \end{subfigure}
        \caption{These figures show the local and global feature explanations for both the NSL-KDD and CIC-IDS datasets. (a)(c) Demonstrates features the GSOM has chosen for a malicous sample from the NSL-KDD and CIC-IDS-2017 datasets. The more siginficant a feature is, the higher its value. (b)(d) Global feature significance is calculated using Bayesian Probability of Significance \cite{Hamel2012a}. Features that have a higher significance value are much more likely to cause a prediction to be made for benign or anomalous. The global explanations apply to all of the tested models}
\label{fig: local-global}
\end{figure*}

Our CL based architecture and its SOM variants were evaluated on both \textit{traditional performative tests} and \textit{explanation generation}. The datasets used to test our architecture were NSL-KDD and CIC-IDS-2017 (see Section \ref{sec:Architectures_Pre-Modeling}). In this section, we examine the performative results from all our CL models, and the explanation results from the Growing Self Organizing Map (GSOM) model. The GSOM explanations were chosen because of the high accuracy of the GSOM model and its similarity to both the SOM and GHSOM.

\subsection{Performative Tests}

\begin{table*}[ht]
\resizebox{\textwidth}{!}{%
\begin{tabular}{lcccccccc}
\hline\hline
\multicolumn{9}{c}{\textbf{NSL-KDD}}                                                                  \\ \hline
\textbf{} &
  \textbf{SOM} &
  \textbf{GSOM} &
  \textbf{GHSOM} &
  \textbf{P-GHSOM} &
  \textbf{\begin{tabular}[c]{@{}c@{}}NDNN \\ Jia et al. \cite{jia2019network}\end{tabular}} &
  \textbf{\begin{tabular}[c]{@{}c@{}}CNN \\ Mohammadpour \\ et al. \cite{mohammadpour2018convolutional}\end{tabular}} &
  \textbf{\begin{tabular}[c]{@{}c@{}}BGRU+MLP \\ Xu et al. \cite{xu2018intrusion}\end{tabular}} &
  \textbf{\begin{tabular}[c]{@{}c@{}}BAT-MC \\ Su et al. \cite{su2020bat} \end{tabular}} \\ \hline
\textbf{Accuracy}             & 90.9\% & 96.7\% & 98.2\% & 98.0\% & 95.0\% & 99.8\% & 99.3\% & 99.2\% \\
\textbf{Precision}            & 97.2\% & 96.6\% & 98.0\% & 98.0\% & -      & -      & -      & -      \\
\textbf{Recall}               & 83.3\% & 96.5\% & 98.3\% & 97.8\% & 97.4\% & -      & 99.3\% & -      \\
\textbf{F1}                   & 89.7\% & 96.6\% & 98.1\% & 97.9\% & 91.4\% & -      & -      & -      \\
\textbf{FPR}                  & 2.2\%  & 3.1\%  & 1.9\%  & 1.8\%  & -      & -      & 0.8\%  & -      \\
\textbf{FNR}                  & 16.6\% & 3.5\%  & 1.6    & 2.2\%  & -      & -      & -      & -      \\
\textbf{Network Size}         & 1      & 1      & 7288   & 574    & -      & -      & -      & -      \\
\textbf{Training Time (s)}    & 8      & 60    & 692    & 816    & -      & -      & -      & -      \\
\textbf{Prediction Time (ms)} & .03    & .03    & .06    & .04    & -      & -      & -      & -      \\ \hline\hline
\multicolumn{9}{c}{\textbf{CIC-IDS-2017}}                                                             \\ \hline
 &
  \textbf{SOM} &
  \textbf{GSOM} &
  \textbf{GHSOM} &
  \textbf{P-GHSOM} &
  \textbf{\begin{tabular}[c]{@{}c@{}}SDCNN\\ Khan et al. \cite{khan2021spectrogram}\end{tabular}} &
  \textbf{\begin{tabular}[c]{@{}c@{}}DNN+RE\\ Almutlaq \\ et al. \cite{almutlaq2022two}\end{tabular}} &
  \textbf{\begin{tabular}[c]{@{}c@{}}SS-Deep-ID\\ Abdel-Basset\\ et al. \cite{abdel2021semi}\end{tabular}} &
  \textbf{\begin{tabular}[c]{@{}c@{}}CNN-IDS*\\ Halbouni \\et al. \cite{halbouni2022cnn}\end{tabular}} \\ \hline
\textbf{Accuracy}             & 79.4\% & 94.6\% & 96.7\% & 95.7\% & 99.3\% & 97.4\% & 99.6\% & 99.6\% \\
\textbf{Precision}            & 83.2\% & 83.7\% & 89.1\% & 86.5\% & 99.1\% & 98.3\% & 99.5\% & 99.7\% \\
\textbf{Recall}               & 42.0\% & 90.0\% & 94.5\% & 92.7\% & 99.7\% & 99.2\% & 99.2\% & 99.4\% \\
\textbf{F1}                   & 55.8\% & 86.7\% & 91.7\% & 89.5\% & 99.4\% & 98.3\% & 99.4\% & 99.7\% \\
\textbf{FPR}                  & 19.0\% & 4.3\%  & 2.8\%  & 3.5\%  & 1.0\%  & -      & 0.7\%  & 0.5\%  \\
\textbf{FNR}                  & 23.0\% & 10.0\% & 5.5\%  & 7.3\%  & 1.0\%  & -      & 0.5\%  & -      \\
\textbf{Network Size}         & 1      & 1      & 16894  & 119    & -      & -      & -      & -      \\
\textbf{Training Time (s)}    & 260    & 1820   & 4299   & 11205  & -    & -      & -      & -      \\
\textbf{Prediction Time (ms)} & .03    & .06    & 1.5    & .03    & - & -      & -      & -      \\
\hline\hline
\end{tabular}%
}
\caption{This table shows the results from testing our CL based X-IDS architecture. We compare our results with existing black box EBL models including deep neural networks, convoluted neural networks, and various ensemble methods.}
\label{results table}
\end{table*}

%This table shows the results from testing our CL based X-IDS architecture. Additionally, we compare our results with existing black box EBL models . Algorithms denoted with an `*' used either no feature selection or different feature selection from what was described in Section \ref{sec:Architectures_Pre-Modeling}. \textbf{NSL-KDD} New Deep Neural Network (NDNN) \cite{jia2019network}, Convoluted Neural Network (CNN) \cite{mohammadpour2018convolutional}, Bidirectional Gated Recurrent Unit (BGRU) + Multi Layered Perceptron (MLP) \cite{xu2018intrusion}, Bidirectional Long Short-term memory Attention Mechanism (BAT-MC) \cite{su2020bat} \textbf{CIC-IDS-2017} Spectrogram-based Deep Convoluted Neural Network (SDCNN) \cite{khan2021spectrogram}, Deep Neural Network + Rule Extraction (DNN+RE) \cite{almutlaq2022two}, Semi-Supervised Spatiotemporal Deep Learning forIntrusions Detection(SS-Deep-ID)\cite{abdel2021semi}, Convoluted Neural Network Intrusion Detection System (CNN-IDS)\cite{halbouni2022cnn}

The experimental results consist of accuracy, precision, recall, f1, false positive rate, false negative rate, and network size measures. Accuracy refers to the percentage of correct predictions compared to the total test size. Precision measures the ratio of true positive predictions to the total number of positive predictions. Recall is the measure of true positive predictions to the total number of positive samples in the test set. The f1 score is a measure that gives equal weight of precision and recall. False positive rate is the rate of false positive predictions compared to the amount of ground truth negatives. False negative rate is the rate of false negative predictions compared to the amount of ground truth positives. Network size is simply the amount of GSOMs within the hierarchical GHSOM structure. For SOMs and GSOMs, the network size is 1. The training time of each algorithm and the average time for a single prediction is also measured. All results can be found in Table \ref{results table}.

%Performative tests allow us to compare our architecture to others in the literature. 
\subsubsection{Testing Parameters}
The parameters selected for our models can be found in Table \ref{table: parameters}. In this work, the SOM was setup to run over 1000 epochs using an 18 $x$ 18 map. We found that increasing the number of epochs served to overfit the models and decrease the overall efficacy of the model. Both the size of the map and the number of training epochs were chosen through trial and error. The GSOM parameters were set to 100 and 40 training epochs for NSL-KDD and CIC-IDS-2017 respectively. We found that this in addition to an aggressive Spread Factor (SF) of .9 created the best performative results. The GHSOM parameters were set to 100 epochs per GSOM created using a SF of .3 and a Learning Rate (LR) of .006. These settings were discovered using the parameter selection process outlined in Section \ref{sec:Architectures_Post-Modeling}. Using these parameters, we were able to create well trained, highly accurate models.

\begin{figure*}[h]
     %\centering
     \begin{subfigure}{0.49\textwidth}
         \centering
         \includegraphics[width=\textwidth]{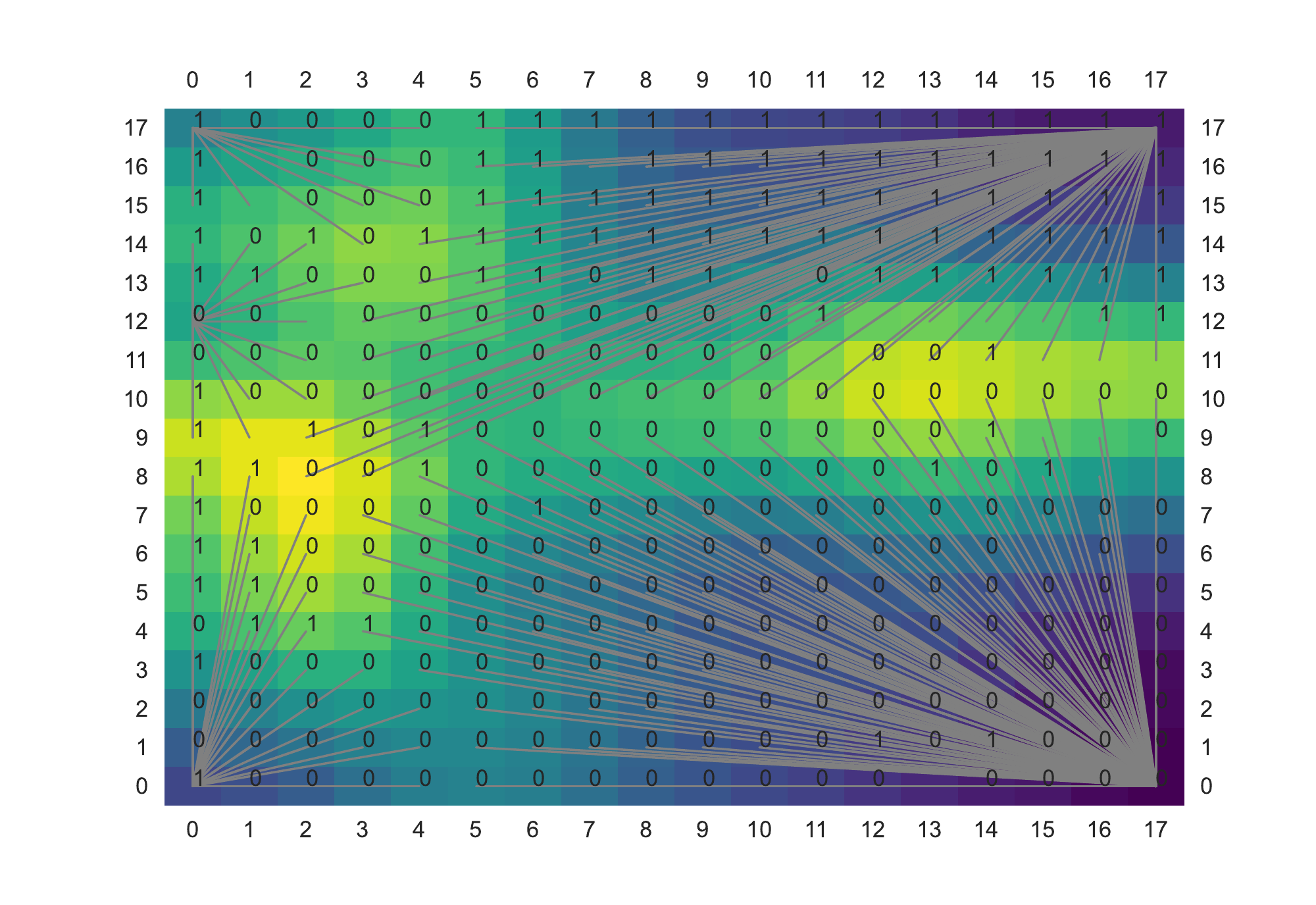}
         \vspace*{-8mm}
         \caption{NSL-KDD Starburst U-Matrix}
         \label{fig: nsl-U-Matrix}
     \end{subfigure}
     \hfill%
     \begin{subfigure}{.34\textwidth}
         \centering
         \includegraphics[width=\textwidth]{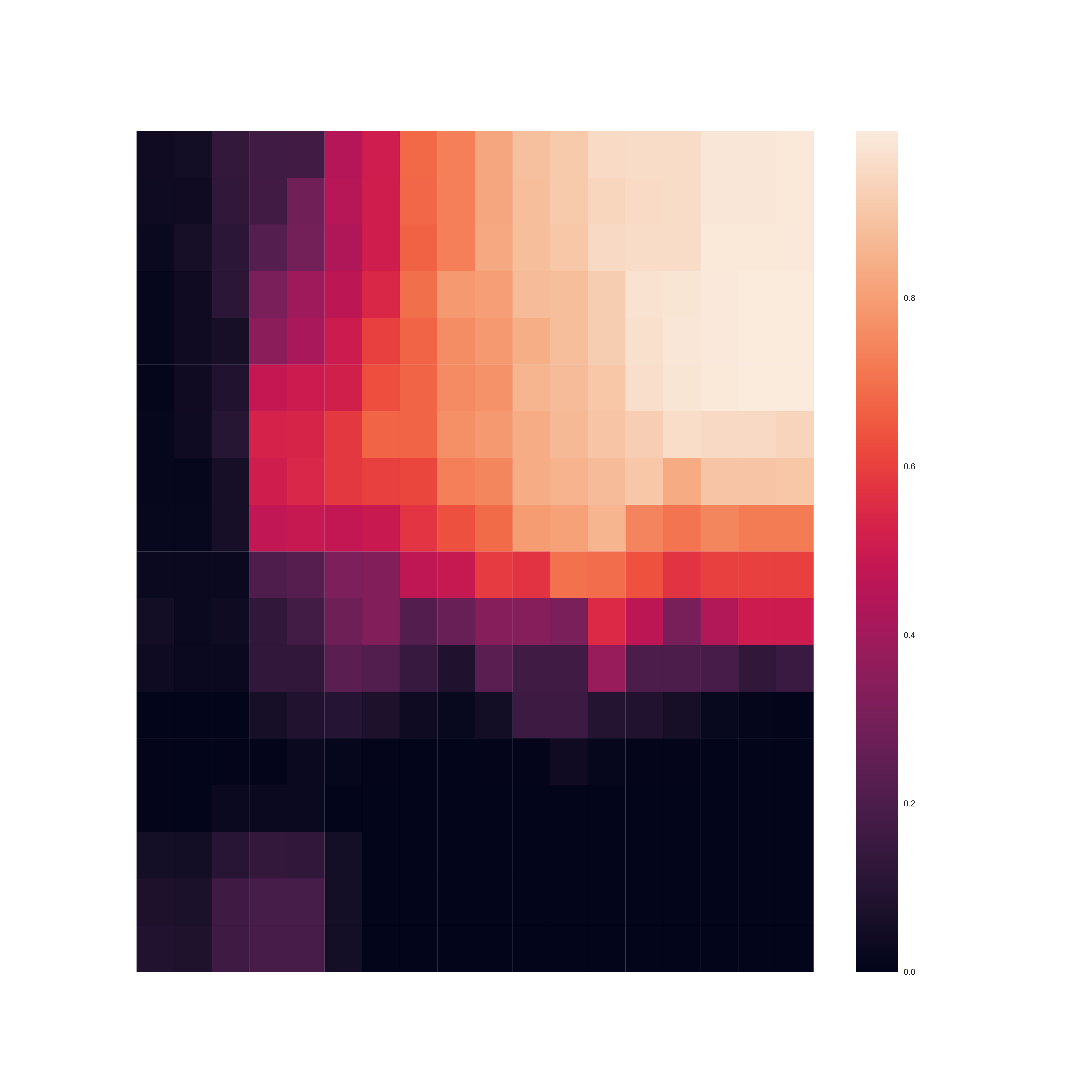}
         \vspace*{-8mm}
         \caption{Dst byte Feature Map}
         \label{fig: nsl-Feature}
     \end{subfigure}
     \hfill%
     \vfill
     \begin{subfigure}{0.49\textwidth}
         %\hspace{40pt}
         %\vspace{2}
         \centering
         \subcaptionbox{CIC-IDS-2017 Starburst U-Matrix\label{fig: cic-U-Matrix}}{%
         \includegraphics[width=\textwidth]{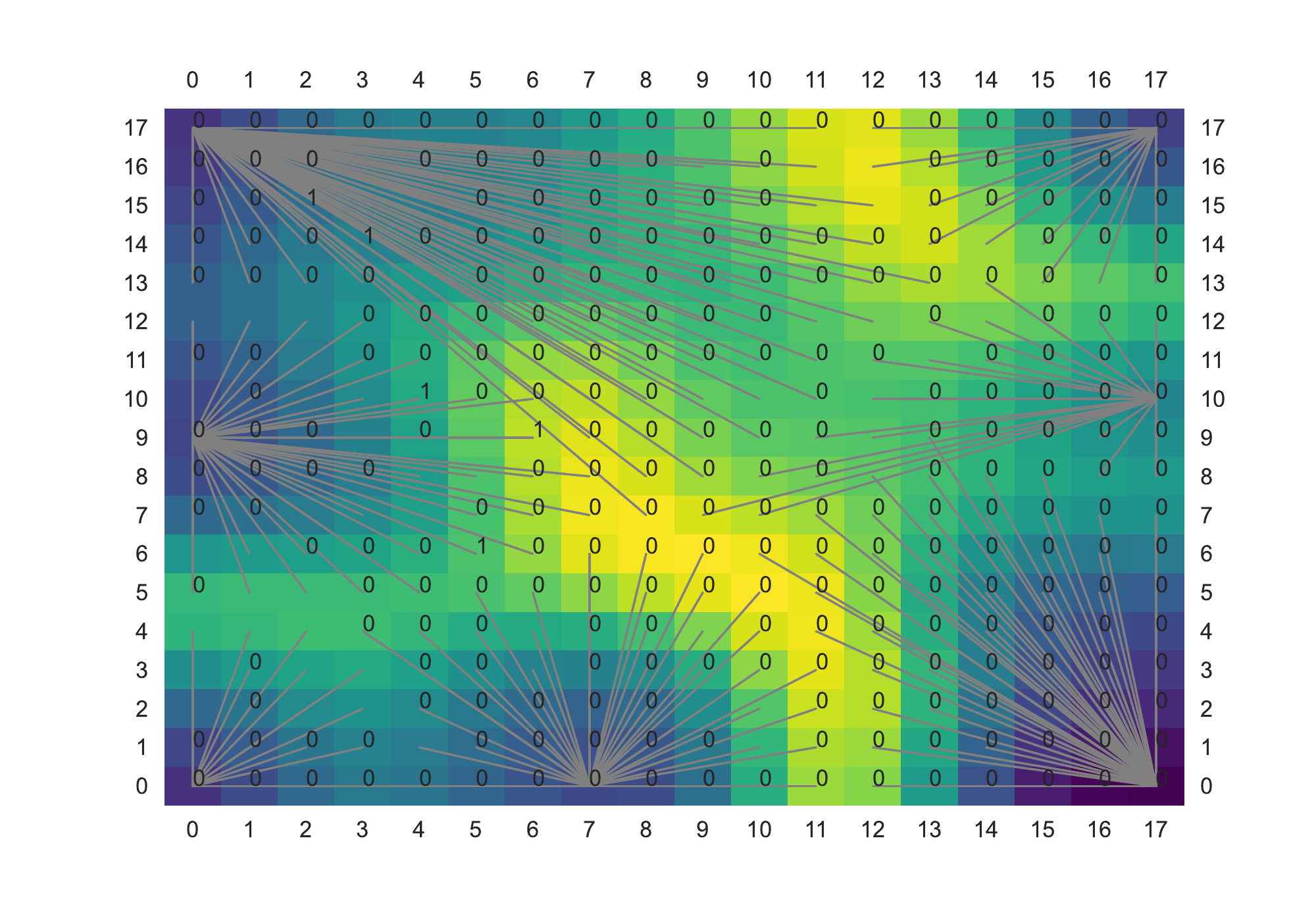}
         \vspace*{-5mm}
         }
         \label{fig: cic-U-matrix}
     \end{subfigure}
     \hfill%
     \begin{subfigure}{.34\textwidth}
        %\hspace{80pt}
        \centering
        \subcaptionbox{Flow bytes/s Feature Map\label{fig: cic-Feature}}{%
        \includegraphics[width=\textwidth]{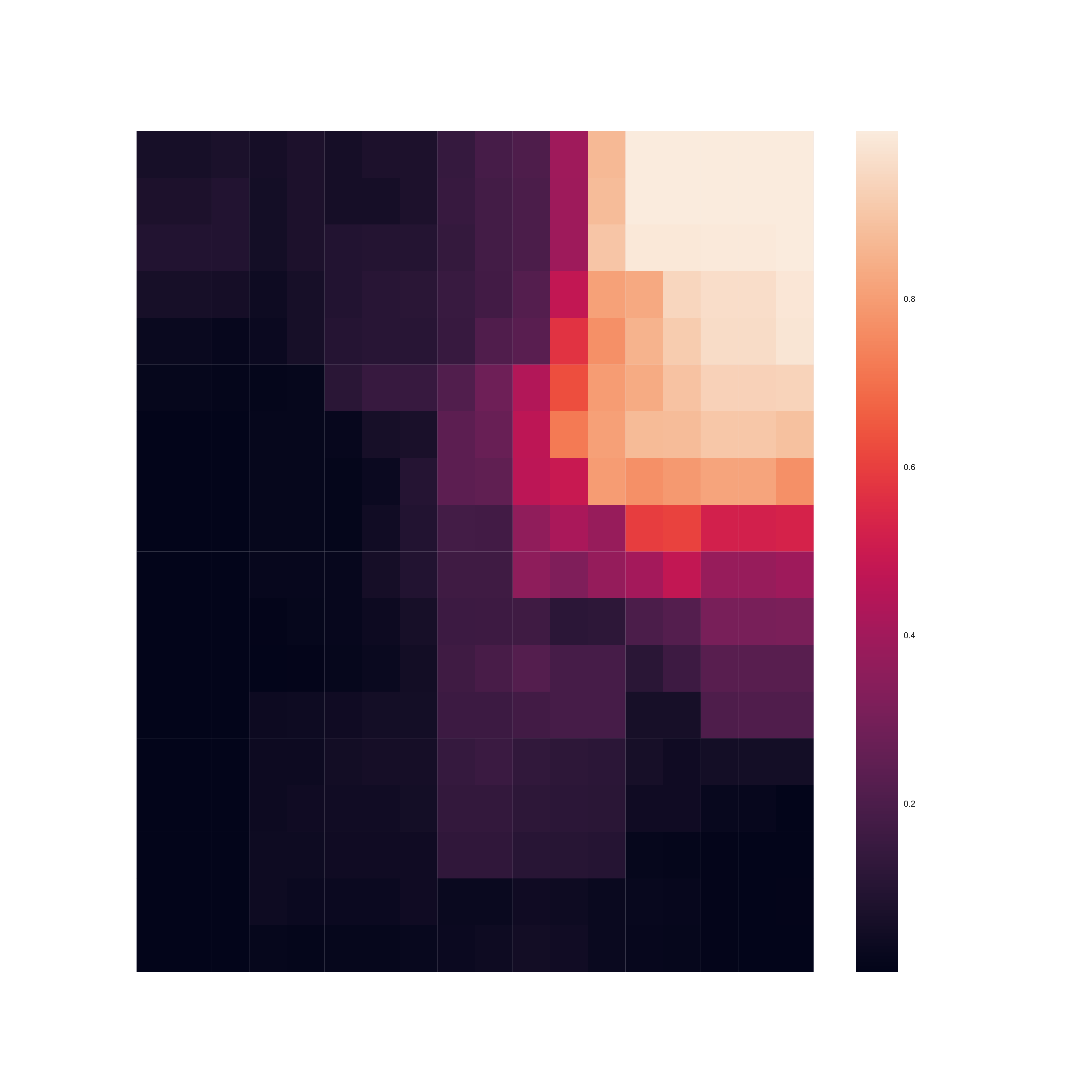}
        \vspace*{-5mm}
        }
        \label{fig: cic-feature}
         
     \end{subfigure}
        \caption{(a)(d)The Starburst U-Matrix shows both the most common label for each node and the clusters the SOM has learned. Darker areas represent units that are close Euclidean Distance-wise. Notably, we can see a clear divide between classes on the NSL-KDD dataset as represented in the figure. (b)(e) K-means clustering can be used as a simplified view of where labels appear on the SOM. In this model's iteration, anomalous traffic is mostly grouped on the bottom of the SOM.(c) The feature value heatmap displays the value of a specific feature on each unit in the SOM. Lighter values represent units with values closer to 1, while darker values show values closer to 0. The `dst byte' example shows that the bottom `anomalous' cluster values higher values. }
        \label{fig: SOM visualizations}
        
\end{figure*}

\subsubsection{Performative Results}

The first part of our experiments looked at how accurate our architecture could be. The SOM performed the worst out of our set of CL algorithms. This is expected as it is the least complex of the three algorithms. It achieved an accuracy of 90.9\% on NSL-KDD and 79.4\% on CIC-IDS-2017. The majority of its accuracy loss comes from its high false negative rate for NSL-KDD and both error rates for CIC-IDS-2017. Out of all of the models, the SOM is the fastest to train and predict with. This may not be a good trade-off, however, with how low its accuracy is. On the other hand, the GSOM, using a more complex algorithm, is able to achieve a much higher accuracy of 96.7\%. The GSOM's growing nature allows it to adapt to new benign and malicious behavior, reducing the number of false negatives. The gain in accuracy comes with the cost of additional training time. Finally, we tested the GHSOM with and without feature selection. The feature selected GHSOM produces an accuracy of 96.2\% and 96.1\% for the normal and pruned variants on the NSL-KDD dataset. The CIC-IDS-2017 feature selected models both achieved accuracies of 96.0\%. When run without feature selection, the model increases its accuracy to 98.2\% and 98.0\% for normal and pruned NSL-KDD variants and 96.7\% and 95.7\% for CIC-IDS-2017. The GHSOM is able to perform better when it has more features. Because of its ability to grow both vertically and horizontally, it is able to make connections in data that its predecessors can not. We also tested the GSOM model using all features. Unlike the GHSOM, it does not benefit from having more information. Overall, the GHSOM with no feature selection performs the best out of all of our CL models.

The models from our architecture can also be compared to other algorithms in the literature. Some of the best performing algorithms are variations of Deep Neural Networks (DNN) and other black box Neural Networks (NN). These black box models are able to capture complexities in data that no human or white box model could comprehend. However, a major problem these black box models have is that they are not easily explainable. Unlike the CL algorithms detailed above, information is given to these NNs and no explanation is given as to why a prediction is made. Table \ref{results table} compares our results to some works in the literature. Note that these models either have no feature selection or a different feature selection than our own models, and the authors did not list all possible metrics. The NSL-KDD and CIC-IDS-2017 dataset models are all similar in accuracy. Our GHSOM architecture has between 1\% - 3\% lower accuracy compared to others, however, the GHSOM is by far more explainable. Even with the small loss in accuracy, we believe that an explainable IDS can be more beneficial for users. Using the explanations given to us by our models, it may be possible to modify parts of the architecture to match the black box models' accuracy.

\subsection{Explanation Generation}
Explanation generation can be divided into two subcategories: Statistical and Visual. Statistical explanations for our architecture consist of Global and local feature significance charts. These explanations can help build a general understanding of the model but lack topographical information. Visual explanations are datamined from the CL models and allow the user to combine the statistical explanations with topographical information. These include the U-matrix, feature heatmap, and label map. In this section, we will discuss the GSOM's explanations. The GSOM performed just as well as the GHSOM with regard to accuracy and uses the same explanations as the GHSOM. However, the GHSOM may require the user to look at its many different hierarchical GSOMs to be useful. Therefore, any method used to understand the GSOM model can also be expanded to the GHSOM.

\subsubsection{Global Explanations}
Global significance explanations allow the user to form a general understanding of models and datasets. `Bayesian probability of significance' is the method we chose to create global explanations \cite{Hamel2012a}. This algorithm calculates feature variance where, in theory, features with higher variance are more likely to cause clustering in data. Using explanations created with this method, users can begin to create a strategy for examining explanations. This can help a user when faced with a dataset with many features. However, a global explanation is not guaranteed to be useful for all predictions. This type of global explanation is probabilistic in nature.  Some local predictions may use the least probable features when choosing a label. Global feature significance explanations should be used as a guide to help examine the datasets and explanatory outputs.

%Investigating datasets with large amounts of features can be daunting when trying to understand the more fine-grained explanations mentioned below. Features with higher significance are \textit{more likely} to be more important with regard to predictions. They are not guaranteed to impact all predictions.

The global explanations for our two datasets can be found in Figures \ref{fig: nsl-global} and \ref{fig: cic-global}. More important features are denoted with higher significance. For the NSL-KDD dataset, we can see the top three most significant features are `Destination (dst) bytes', `dst host count', and `Source (src) bytes'. Additionally, we can see that even though these three are the most important, the other features have high enough significance that they could play a role in predictions. The CIC-IDS-2017 global explanation tells a different story. `Flow bytes/s' has the highest variance among all of the features. We can classify this as the most significant feature while viewing the next 11 features (`flow duration' to `Forward (fwd) packets/s') as being somewhat significant. The final five features are less likely to impact the model.

Using the global explanations in Figures \ref{fig: nsl-global} and \ref{fig: cic-global}, we can form some initial conclusions about the models and datasets. For NSL-KDD, benign and malicious packets differ in size, length, and number. This is demonstrated by `dst host count', `duration', and `dst bytes'. At this stage, forming any other conclusions may not be beneficial. One of these features may be a good starting point to look at for later explanation types. Malicious and benign traffic in the CIC-IDS-2017 dataset appears to vary in how much data is sent over the network. Similar to the NSL-KDD dataset, it may not be a good idea to form a more concrete opinion about the models just yet. However, we do know that malicious and benign data differ in the ways mentioned above, and these ideas may be good concepts to investigate.

\subsubsection{Local Explanation}
The local prediction explanations for GSOM anomalies can be found in Figures \ref{fig: nsl-local} and \ref{fig: cic-local}. Using local explanations, users can take their coarse understanding of the dataset and begin fine tuning it. Here, we will look at an anomalous local explanation from each dataset, however, it may be beneficial for a user to view many at one time. Local explanations are created by examining a feature's proximity to its BMU counterpart (see Section \ref{sec:Architectures_Post-Modeling}). The most important features have higher values.

%Viewing many local predictions about a specific label can lead a user to understand why a label is chosen. Thus, the user can use global and local explanations to create a mental model of the actual model.

The anomalous prediction for NSL-KDD in Figure \ref{fig: nsl-local} can be used to demonstrate how to understand local prediction explanations. For this anomaly, `Destination (dst) host count', `duration', and 'dst bytes' were the most important features. CIC-IDS-2017's local anomalous explanation has some interesting features. Eight features have a proximity of one or very near to one. Five of its features are of some significance to the prediction, while the first four are less or not significant. 

To be able to better understand how the model labels predictions, multiple local explanations should be used. Ideally, the user would need to look at many anomalous and benign explanations. Doing so would allow the user to solidify their idea of how the model makes a malicious or benign prediction. Looking at our NSL-KDD anomaly explanation, it is similar to the global explanation. `Destination (dst) host count', `dst bytes', and `src bytes' are all of higher importance. However, `duration' has also made a large impact even though it was considered less significant globally. If we look at other malicious prediction explanations, we see a similar trend. The NSL-KDD model uses these features to make many anomalous predictions. The CIC-IDS-2017 explanation states that `flow bytes/s' and seven other features are what cause this anomaly. Similar to NSL-KDD, this trend tends to hold true over many other anomalous predictions. One or more of the features may change places of importance, but a pattern can be seen in the explanations.

The user, now having examined both global and local explanations, has hopefully begun to form a mental model of how the GSOM makes predictions. The process used in this section can be used to form a general understanding of the SOM and GHSOM models. Using clues discovered in their investigation, a user may want to view visual explanations for specific features. The topological information gleaned from these explanations may prove fruitful in understanding the model even better.

\subsubsection{Visual Explanations}

\begin{figure*}[htbp]
    \centering
    \begin{subfigure}[b]{0.285\textwidth}
        \includegraphics[width=\textwidth]{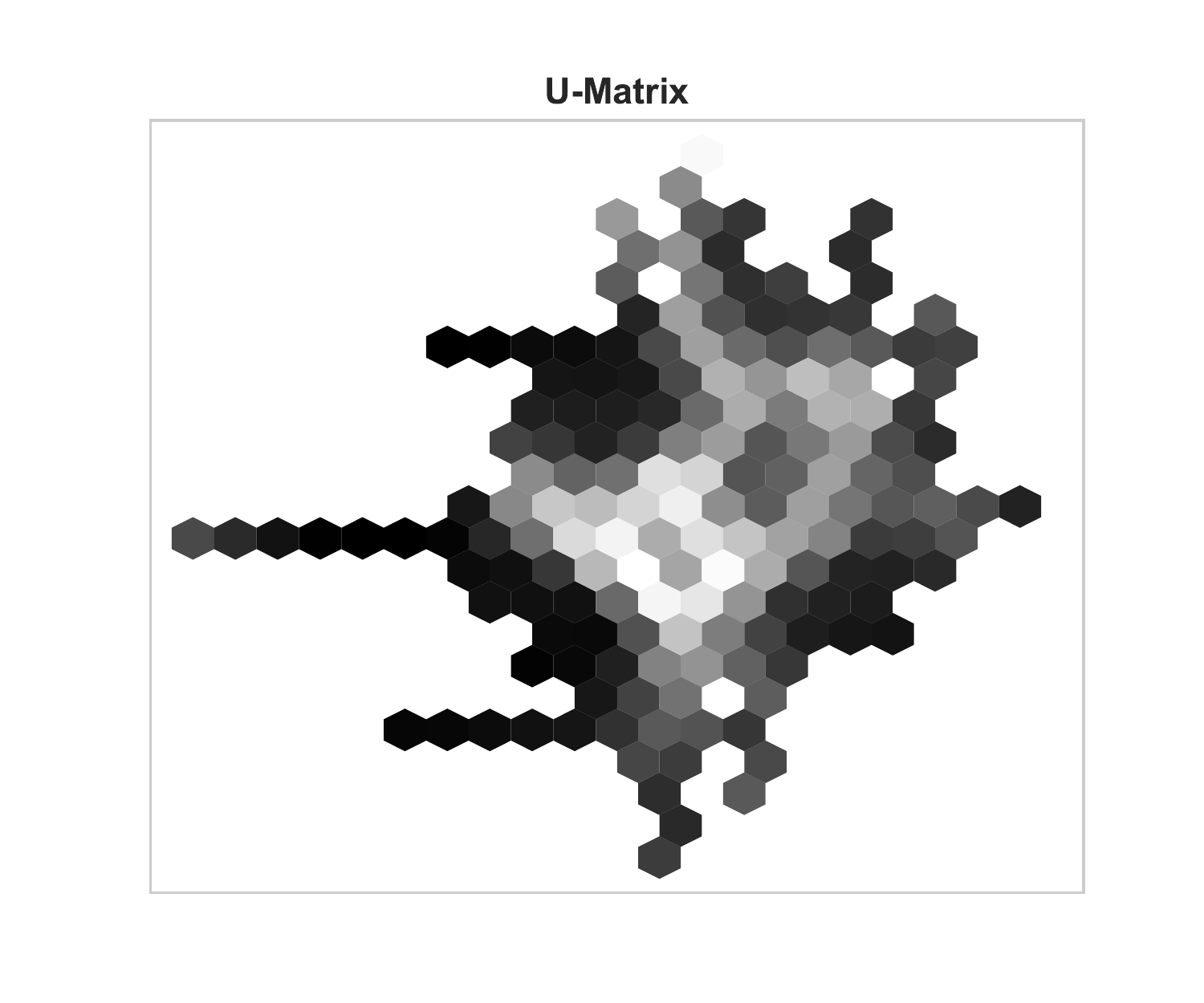}
        \caption{NSL-KDD U-Matrix}
        \label{fig: GSOM-U nsl-kdd}
    \end{subfigure}
    \begin{subfigure}[b]{0.33\textwidth}
        \includegraphics[width=\textwidth]{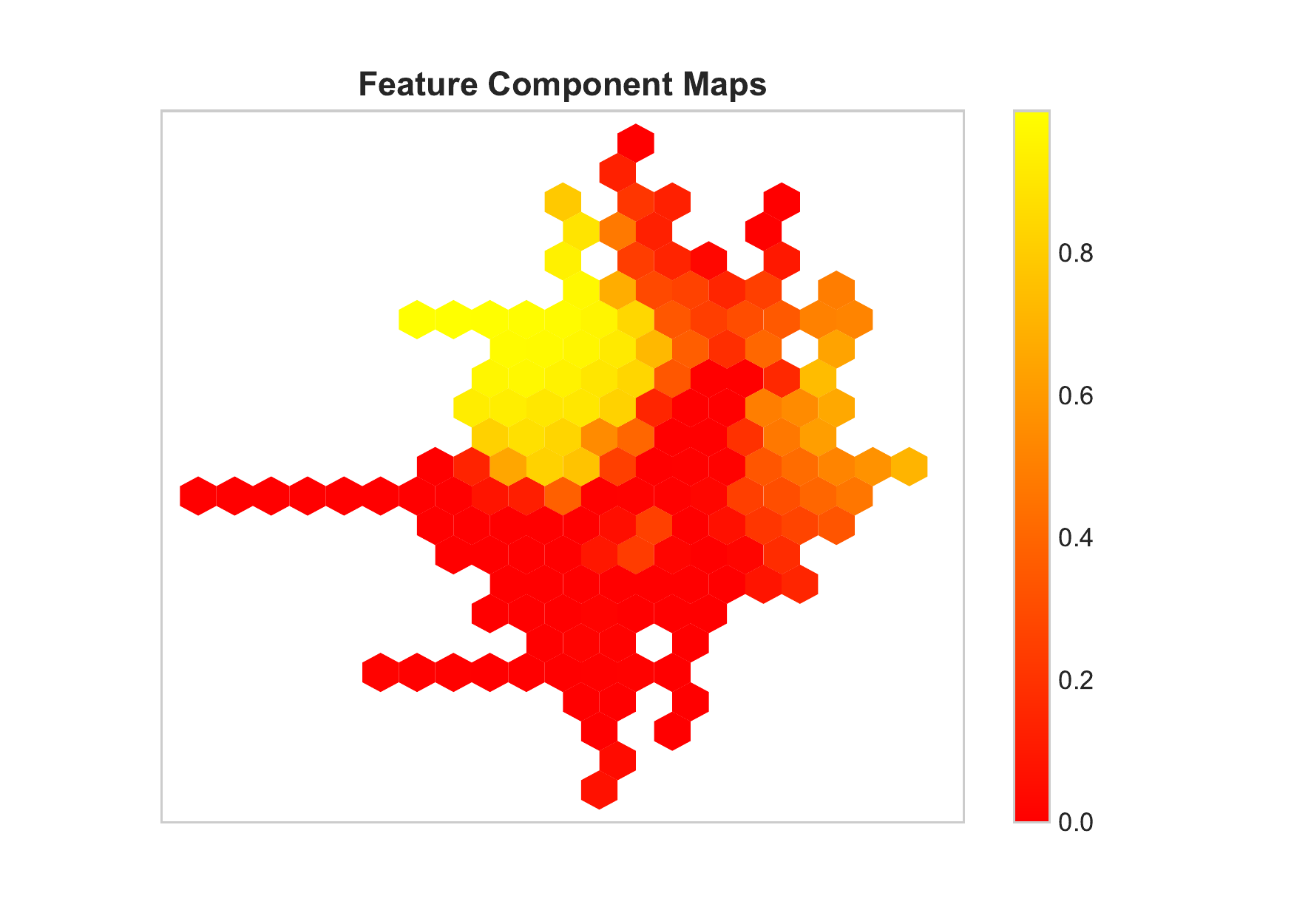}
        \caption{Feature Component Map of Dst-Bytes}
        \label{fig: GSOM-Feature nsl-kdd}
    \end{subfigure}
    \begin{subfigure}[b]{0.33\textwidth}
        \includegraphics[width=\textwidth]{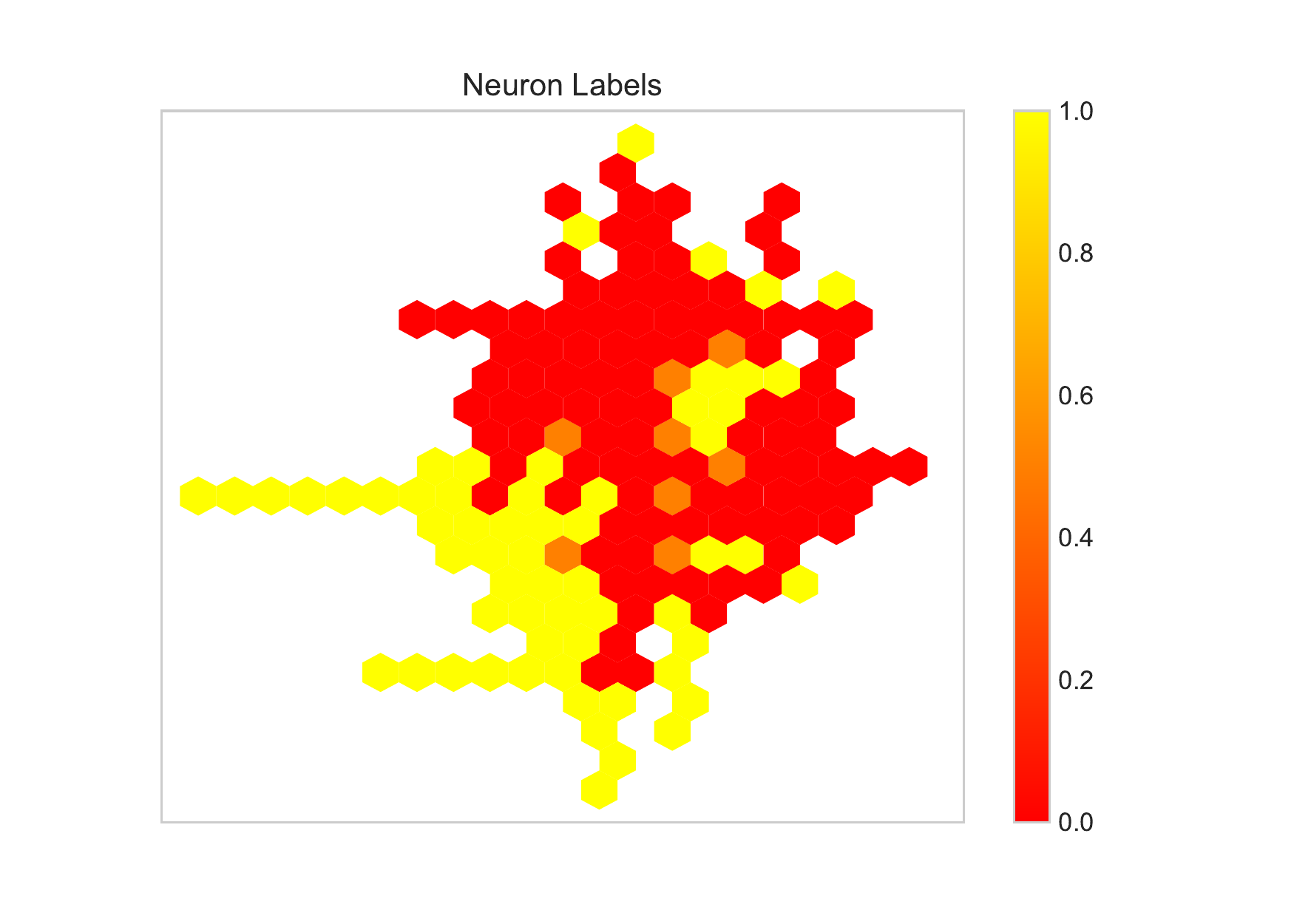}
        \caption{NSL-KDD Neuron Labels}
        \label{fig: GSOM-Labels nsl-kdd}
    \end{subfigure}

    \begin{subfigure}[b]{0.285\textwidth}
        \includegraphics[width=\textwidth]{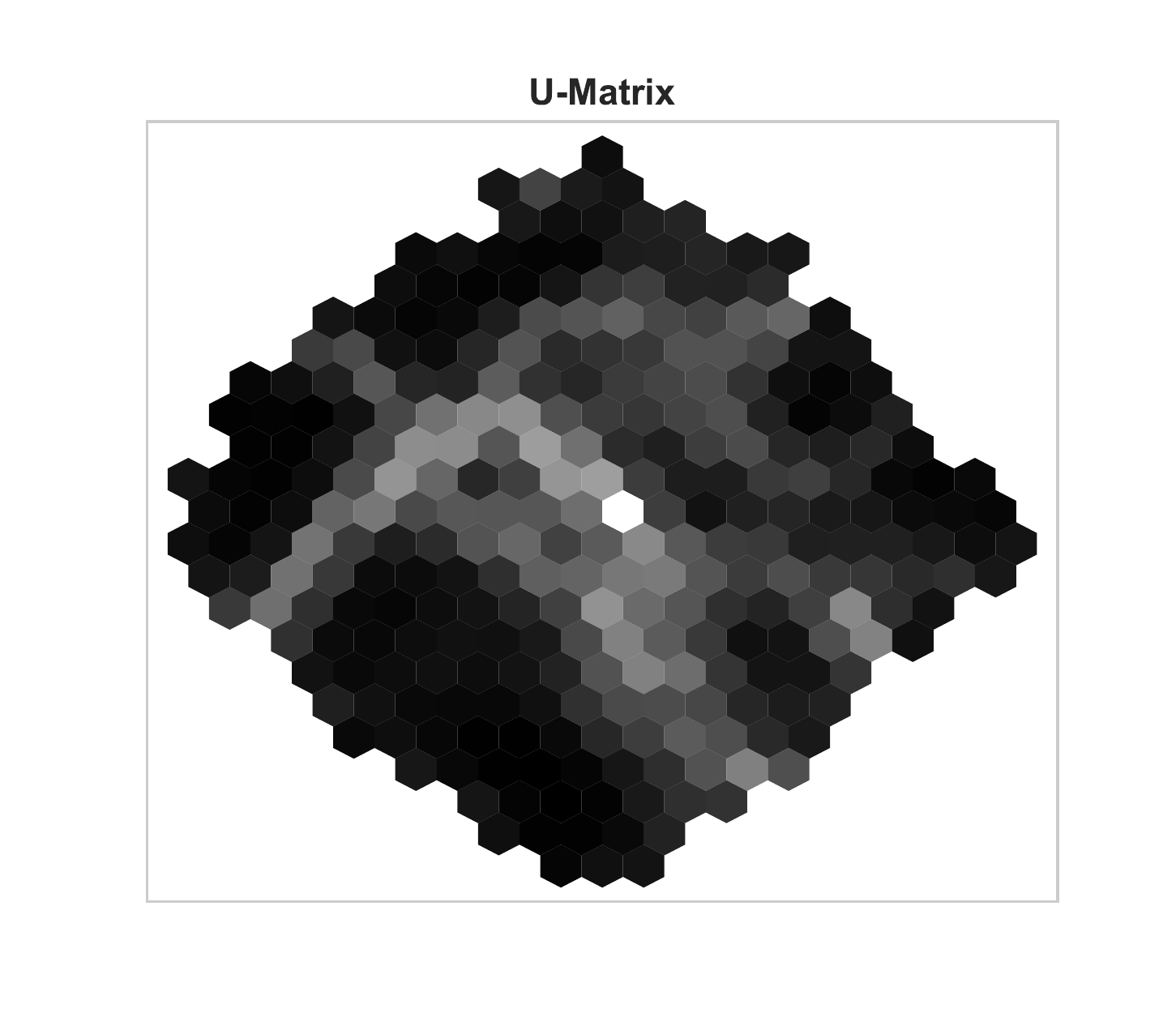}
        \caption{CIC-IDS-2017 U-Matrix}
        \label{fig: GSOM-U cic-ids}
    \end{subfigure}
    \begin{subfigure}[b]{0.33\textwidth}
        \includegraphics[width=\textwidth]{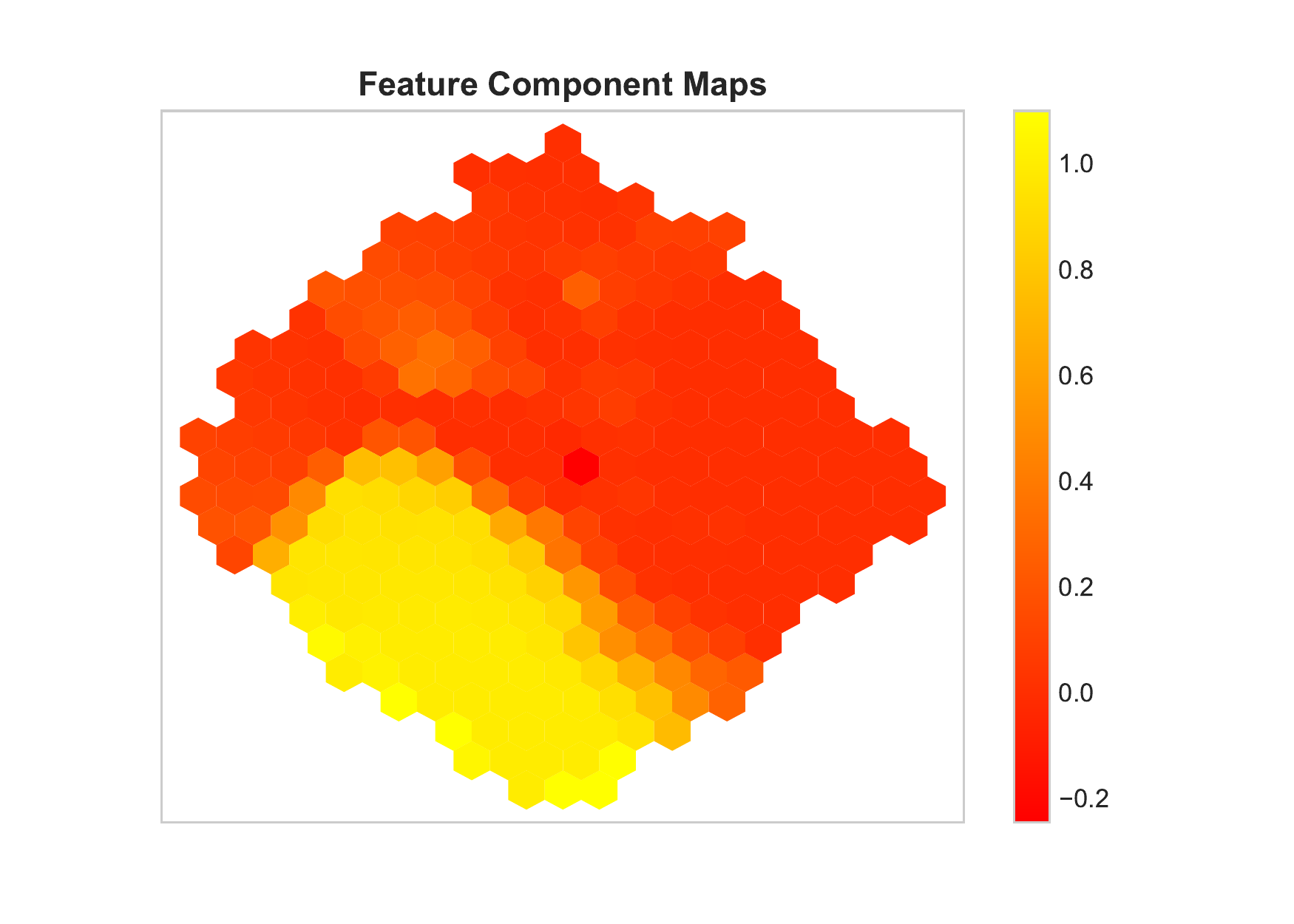}
        \caption{Feature Component Map of Flow Bytes/s}
        \label{fig: GSOM-Feature cic-ids}
    \end{subfigure}
    \begin{subfigure}[b]{0.33\textwidth}
        \includegraphics[width=\textwidth]{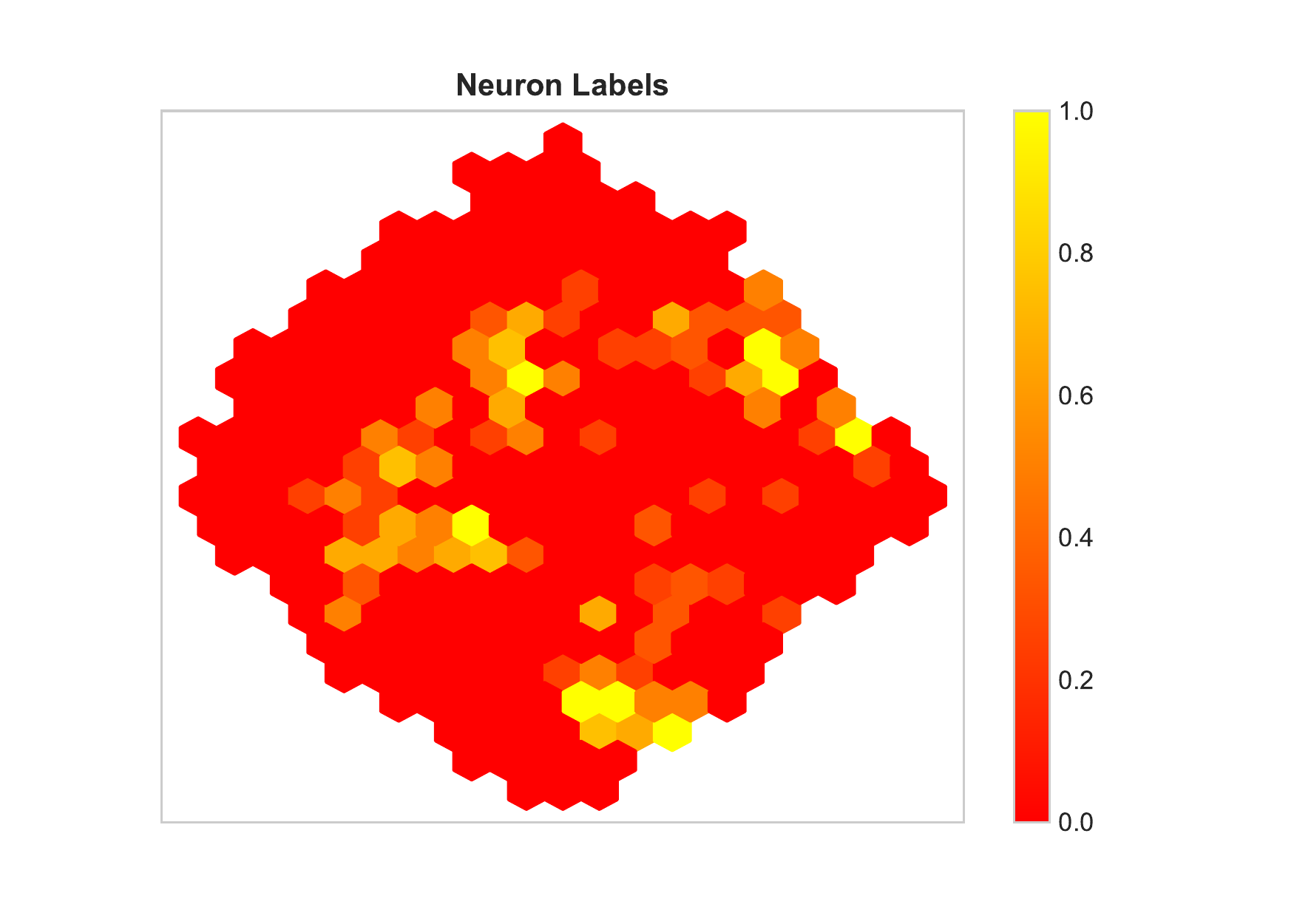}
        \caption{CIC-IDS-2017 Neuron Labels}
        \label{fig: GSOM-Labels cic-ids}
    \end{subfigure}
    \caption{Visualizations generated from a GSOM for models trained on NSL-KDD and CIC-IDS-2107. (a)(d) The U-matrix maintains the same properties as the SOM starburst visualization with darker areas representing neurons closer together. (b)(e) The Feature Component Map also shares the same properties as the SOM feature map in Figure 5.2. (c)(f) The Neuron Label map shows the class label represented by a red or yellow color.}
    \label{GSOM Visualizations}
\end{figure*}

\begin{comment}
\begin{figure*}[htbp]
    \centering
    \begin{subfigure}[b]{0.285\textwidth}
        \includegraphics[width=\textwidth]{figures/SOM figs/U_cic.pdf}
        \caption{U-Matrix}
        \label{GSOM_U}
    \end{subfigure}
    \begin{subfigure}[b]{0.33\textwidth}
        \includegraphics[width=\textwidth]{figures/SOM figs/Features_cic.pdf}
        \caption{Feature Component Map of Flow Bytes/s}
        \label{GSOM_feature}
    \end{subfigure}
    \begin{subfigure}[b]{0.33\textwidth}
        \includegraphics[width=\textwidth]{figures/SOM figs/Labels_cic.pdf}
        \caption{Neuron Labels}
        \label{GSOM_label}
    \end{subfigure}
    \caption{Visualizations generated from a DBGSOM trained on CIC-IDS-2017. (a)(b)(c) Share the same properties with Figure \ref{GSOM Visualizations}}
    \label{GSOM-Vis-cic}
\end{figure*}
\end{comment}

Unlike the statistical explanations, the three visual explanations need to be viewed together. Examples of these explanations can be seen in Figure \ref{GSOM Visualizations}. The U-matrix is a hexagonal grid composed of dark and light nodes. Darker cells represent nodes in the GSOM that are closer to one another. Lighter cells denote a separation of nodes and clusters. The label map contains the label used by each node for prediction. Light yellow nodes assign a prediction label of \textit{malicious} while dark red nodes assign a prediction label of \textit{benign}. A shade in between yellow and red indicate when a node has a probability of choosing either \textit{benign} or \textit{malicious}. Finally, the feature heatmap contains visualized information of a single feature in the GSOM. Since each node represents a sample from the dataset, we are able to visually map feature values. Lighter colored nodes contain a feature value closer to one, while the darker values are closer to zero.

Firstly, we can look at the U-matrix to see how many clusters there are and where they formed. The NSL-KDD model created about five clusters seen in Figure \ref{fig: GSOM-U nsl-kdd}. Four clusters appear on the edge of the map with a smaller one in the middle. In Figure \ref{fig: GSOM-U cic-ids}, the CIC-IDS-2017 model created between four to six clusters. Similar to the NSL-KDD U-matrix, the user may see that there is a cluster at the top, left, bottom, right, and middle of the map. Now that we have defined the clusters, we can look at each cluster's associated label. The NSL-KDD model appears to label the bottom-left and middle clusters as anomalous. The other clusters seem to be mostly benign. For CIC-IDS-2017, we can see that nearly the entire map is benign. There are a few clusters of malicious labels, but they are intermixed with benign data. Lastly, we can look at the meaning of the feature heat map. The NSL-KDD feature heat map represents the `desetination (dst) bytes' feature. Here, we can see that the top-left of the map contains higher values of `dst bytes' while the rest of the map contains much lower values. Similarly, the CIC-IDS-2017 dataset's `flow bytes/s' is much higher in the bottom-left than anywhere else on the map.

With our new understanding of visual explanations, we can combine them to form more complex ideas about the model. A user can view both the feature heat map and the U-matrix and see that there is a cluster of nodes associated with high `dst bytes' values. They can then look to the neuron label map and see that this cluster is associated with benign data. This thought process can be used with the other features from the dataset. The user can then decide that they want to view the dataset's next most significant feature. In NSL-KDD's case, the next best feature would be `dst host count'. Alternatively, the user could decide to use a local explanation to make decisions about which features to view. The process outlined above can be used for larger datasets such as CIC-IDS-2017 to guide the user into forming conclusions about the model.

\subsubsection{User Conclusions}
Having looked at all of the explanations, a user may now make a few decisions. The first is that the explanations were sufficient such that they can complete their \textit{task}. The second is that they believe that the model could perform better or output better explanations that could benefit their \textit{task}. The NSL-KDD explanations may be sufficient for a user, however, the CIC-IDS-2017 model has some flaws that could affect its accuracy. The user could see that the CIC-IDS-2017 explanations heavily favor benign data. Using the architecture diagram defined in Figure \ref{fig:cl_arch}, the user may modify different aspects of the architecture. In this case, the user may decide to preprocess the CIC-IDS-2017 dataset differently. It is possible that stratifying the training portion of the dataset would create a more balanced set of clusters.

\subsubsection{SOM and GHSOM Explanations}
The process outlined above can also be used to understand both SOM and GHSOM explanations. In fact, the global and local explanations for these two algorithms look the same. However, the SOM algorithm that we chose uses a visual explanation that combines the U-matrix and label map. It also includes a starburst-like pattern that helps dictate where the centers of the clusters are and how far they stretch. We leave the SOM explanations in Figure \ref{fig: SOM visualizations} as an exercise for the reader to form their own conclusions using the above process. Lastly, the GHSOM uses the same visual explanations as the GSOM, but it would require many different images of each to explain. Instead, we have created a visualization representing the hierarchical structure of the GHSOM that includes how each GSOM within labels data. This visualization for the pruned GHSOM can be seen in Figures \ref{fig: treemap pruned}. In the figure, red represents a node that labels data as malicious, blue nodes are benign, and yellow indicates a branching node.

\section{Conclusion}
\label{sec:Conclusion}

In this paper, we created an Explainable Intrusion Detection (X-IDS) architecture featuring three Competitive Learning (CL) based algorithms. It was built using DARPA's recommended guidelines for an explainable system. The architecture consists of four phases: Pre-Modeling, Modeling, Post-Modeling Optimization, and Prediction Explanation. In the Pre-Modeling phase, we preprocess datasets and select our initial model parameters. In the Modeling phase, we train the Self Organizing Map (SOM), Growing Self Organizing Map (GSOM) and Growing Hierarchical Self Organizing Map (GHSOM) and record quality metrics. In the Post-Modeling Optimization phase, we find better model parameters to achieve higher accuracy results, and we implement a pruning process for the GHSOM model. Lastly, we generate explanations and allow the user to make modifications to the architecture in the Prediction Explanation phase. When compared with existing Error Based Learning (EBL) algorithms, CL algorithms are less accurate. However, CL algorithms are far more explainable, leading to a more trustworthy IDS.

The main objective of this paper was to demonstrate the explanatory properties of CL algorithms. In our explanation discussion, we showed that CL algorithms are highly explainable because of their ability to mimic patterns in data. This is a feature that EBL techniques lack. We demonstrated a strategy that a user could use to be able to understand explanations to better trust or improve the model. This strategy involved using course grain explanations such as the global and local feature significance charts to form a general understanding of the models. With this general understanding, users can then use the feature heatmaps, U-matrices, and label maps to form a more comprehensive knowledge of the model.

Additionally, a pruning process was applied on the GHSOM in an effort to lower the number of branches it generated. We were able to decrease the size of the GHSOM by 92\% - 99\% while only losing 0.2\% - 1.0\% accuracy. This lowers the performative overhead and allows the pruned GHSOM to make predictions faster than the unpruned GHSOM. Additionally, reducing the size of the GHSOM can help with visualizing explanations.

Lastly, a performance analysis was performed on our CL-based X-IDS architecture. Tests were run using the NSL-KDD and CIC-IDS-2017 datasets. Our experimental results showed that the CL models can achieve accuracies as high as 98.2\% on the NSL-KDD dataset and 96.7\% on the CIC-IDS-2017 dataset. We compare these results with existing EBL algorithms. We find that EBL algorithms are 1\% to 3\% more accurate than the CL algorithms, however, EBL models are far less explainable. 

%Future works for our research include investigating ways to make the GHSOM more explaianble. The GHSOM currently suffers from having too many GSOMs. A user would potentially need to look at hundreds of different GSOMs to be able to understand the model on a global scale. Local predictions do not suffer from this as the GSOM the prediction comes from could be outputed along with all previous explanations from this paper. A potential solution to this problem is to create a system that summarizes branches. This may require the model to be trained using a multi-class intrusion detection datasets so that the explanations can be more direct.

The future for intrusion detection is explainability. Using architectures and methods, such as the ones used in this paper, will lead to more powerful and trustworthy IDS. White box methods can be improved and adapted to create more accurate AI models. Competitive Learning algorithms embody this philosophy. Their explainability, ease of use, and low performative cost allow for them to be the front runners for future X-IDS.

\section*{Acknowledgement}

\noindent This work by Mississippi State University was financially supported by the U.S. Department of Defense (DoD) High Performance Computing Modernization Program, through the US Army Engineering Research and Develop Center (ERDC) (\#W912HZ-21-C0058). The views and conclusions contained herein are those of the authors and should not be interpreted as necessarily representing the official policies or endorsements, either expressed or implied, of the U.S. Army ERDC or the U.S. DoD.

%The authors of this paper would like to acknowledge Michael Allen's work to help improve the paper's overall readability.

%\appendix

%\section{Sample Appendix Section}

%\input{figures/tex_files/ghsom_branch_visualization.tex}

 \bibliographystyle{elsarticle-num}%[nosrt] 
 \bibliography{cas-refs}

\begin{thebibliography}{10}
\expandafter\ifx\csname url\endcsname\relax
  \def\url#1{\texttt{#1}}\fi
\expandafter\ifx\csname urlprefix\endcsname\relax\def\urlprefix{URL }\fi
\expandafter\ifx\csname href\endcsname\relax
  \def\href#1#2{#2} \def\path#1{#1}\fi

\bibitem{neupane2022explainable}
S.~Neupane, J.~Ables, W.~Anderson, S.~Mittal, S.~Rahimi, I.~Banicescu,
  M.~Seale, Explainable intrusion detection systems (x-ids): A survey of
  current methods, challenges, and opportunities, arXiv preprint
  arXiv:2207.06236 (2022).

\bibitem{iannucci2021performance}
S.~Iannucci, J.~Ables, W.~Anderson, B.~Abburi, V.~Cardellini, I.~Banicescu, A
  performance-oriented comparison of neural network approaches for
  anomaly-based intrusion detection, in: 2021 IEEE Symposium Series on
  Computational Intelligence (SSCI), IEEE, 2021, pp. 1--7.

\bibitem{buczak2015survey}
A.~L. Buczak, E.~Guven, A survey of data mining and machine learning methods
  for cyber security intrusion detection, IEEE Communications surveys \&
  tutorials 18~(2) (2015) 1153--1176.

\bibitem{Marshan}
A.~Marshan, Artificial intelligence: Explainability, ethical issues and bias,
  Annals of Robotics and Automation (2021) 034--037\href
  {https://doi.org/10.17352/ara.000011} {\path{doi:10.17352/ara.000011}}.

\bibitem{ables2022creating}
J.~Ables, T.~Kirby, W.~Anderson, S.~Mittal, S.~Rahimi, I.~Banicescu, M.~Seale,
  {Creating an Explainable Intrusion Detection System Using Self Organizing
  Maps}, in: IEEE Symposium on Computational Intelligence in Cyber Security,
  2022.

\bibitem{gunning2019darpa}
D.~Gunning, D.~Aha, Darpa’s explainable artificial intelligence (xai)
  program, AI Magazine 40~(2) (2019) 44--58.

\bibitem{ribeiro2016should}
M.~T. Ribeiro, S.~Singh, C.~Guestrin, "{W}hy should i trust you?" explaining
  the predictions of any classifier, in: Proceedings of the 22nd ACM SIGKDD
  international conference on knowledge discovery and data mining, 2016, pp.
  1135--1144.

\bibitem{lundberg2017unified}
S.~M. Lundberg, S.-I. Lee, A unified approach to interpreting model
  predictions, Advances in neural information processing systems 30 (2017).

\bibitem{binder2016layer}
A.~Binder, G.~Montavon, S.~Lapuschkin, K.-R. M{\"u}ller, W.~Samek, Layer-wise
  relevance propagation for neural networks with local renormalization layers,
  in: International Conference on Artificial Neural Networks, Springer, 2016,
  pp. 63--71.

\bibitem{kohonen1982self}
T.~Kohonen, Self-organized formation of topologically correct feature maps,
  Biological cybernetics 43~(1) (1982) 59--69.

\bibitem{fritzke1995growing}
B.~Fritzke, Growing grid—a self-organizing network with constant neighborhood
  range and adaptation strength, Neural processing letters 2~(5) (1995) 9--13.

\bibitem{dittenbach2000growing}
M.~Dittenbach, D.~Merkl, A.~Rauber, The growing hierarchical self-organizing
  map, in: Proceedings of the IEEE-INNS-ENNS International Joint Conference on
  Neural Networks. IJCNN 2000. Neural Computing: New Challenges and
  Perspectives for the New Millennium, Vol.~6, IEEE, 2000, pp. 15--19.

\bibitem{denning1987intrusion}
D.~E. Denning, An intrusion-detection model, IEEE Transactions on software
  engineering~(2) (1987) 222--232.

\bibitem{bace2001intrusion}
R.~G. Bace, P.~Mell, et~al., Intrusion detection systems (2001).

\bibitem{mcdole2021deep}
A.~McDole, M.~Gupta, M.~Abdelsalam, S.~Mittal, M.~Alazab, Deep learning
  techniques for behavioural malware analysis in cloud iaas, in: Malware
  Analysis using Artificial Intelligence and Deep Learning, Springer, 2021.

\bibitem{sharma2014evolution}
A.~Sharma, S.~K. Sahay, Evolution and detection of polymorphic and metamorphic
  malwares: A survey, arXiv preprint arXiv:1406.7061 (2014).

\bibitem{Chandola2009AnomalyDA}
V.~Chandola, A.~Banerjee, V.~Kumar, Anomaly detection: A survey, ACM Comput.
  Surv. 41 (2009) 15:1--15:58.

\bibitem{mcdole2020analyzing}
A.~McDole, M.~Abdelsalam, M.~Gupta, S.~Mittal, Analyzing cnn based behavioural
  malware detection techniques on cloud iaas, in: International Conference on
  Cloud Computing, Springer, 2020, pp. 64--79.

\bibitem{szczepanski2020achieving}
M.~Szczepa{\'n}ski, M.~Chora{\'s}, M.~Pawlicki, R.~Kozik, Achieving
  explainability of intrusion detection system by hybrid oracle-explainer
  approach, in: 2020 International Joint Conference on Neural Networks (IJCNN),
  IEEE, 2020, pp. 1--8.

\bibitem{pang2021explainable}
G.~Pang, C.~Ding, C.~Shen, A.~v.~d. Hengel, Explainable deep few-shot anomaly
  detection with deviation networks, arXiv preprint arXiv:2108.00462 (2021).

\bibitem{subba2015intrusion}
B.~Subba, S.~Biswas, S.~Karmakar, Intrusion detection systems using linear
  discriminant analysis and logistic regression, in: 2015 Annual IEEE India
  Conference (INDICON), IEEE, 2015, pp. 1--6.

\bibitem{mahbooba2021explainable}
B.~Mahbooba, M.~Timilsina, R.~Sahal, M.~Serrano, Explainable artificial
  intelligence (xai) to enhance trust management in intrusion detection systems
  using decision tree model, Complexity 2021 (2021).

\bibitem{langin2011annabell}
C.~Langin, M.~Wainer, S.~Rahimi, Annabell island: a 3d color hexagonal som for
  visual intrusion detection, Internation Journal of Computer Science and
  Information Security 9~(1) (2011) 1--7.

\bibitem{Liu2008IsolationF}
F.~T. Liu, K.~M. Ting, Z.-H. Zhou, Isolation forest, 2008 Eighth IEEE
  International Conference on Data Mining (2008) 413--422.

\bibitem{Schlkopf1999SupportVM}
B.~Sch{\"o}lkopf, R.~C. Williamson, A.~Smola, J.~Shawe-Taylor, J.~C. Platt,
  Support vector method for novelty detection, in: NIPS, 1999.

\bibitem{ZhangNN}
G.~Zhang, Neural networks for classification: a survey, IEEE Transactions on
  Systems, Man, and Cybernetics, Part C (Applications and Reviews) 30~(4)
  (2000) 451--462.
\newblock \href {https://doi.org/10.1109/5326.897072}
  {\path{doi:10.1109/5326.897072}}.

\bibitem{moore1988explanation}
J.~D. Moore, W.~R. Swartout, Explanation in expert systemss: A survey, Tech.
  rep., University of Southern California Marina Del Rey Information Sciences
  Inst (1988).

\bibitem{shortliffe1974mycin}
E.~H. Shortliffe, Mycin: a rule-based computer program for advising physicians
  regarding antimicrobial therapy selection., Tech. rep., Stanford Univ Calif
  Dept of Computer Science (1974).

\bibitem{holzinger2017we}
A.~Holzinger, C.~Biemann, C.~S. Pattichis, D.~B. Kell, What do we need to build
  explainable ai systems for the medical domain?, arXiv preprint
  arXiv:1712.09923 (2017).

\bibitem{lindsay2020explainable}
L.~Lindsay, S.~Coleman, D.~Kerr, B.~Taylor, A.~Moorhead, Explainable artificial
  intelligence for falls prediction, in: International Conference on Advances
  in Computing and Data Sciences, Springer, 2020, pp. 76--84.

\bibitem{chun2021study}
Y.~E. Chun, S.~B. Kim, J.~Y. Lee, J.~H. Woo, Study on credit rating model using
  explainable ai, The Korean Data \& Information Science Society 32~(2) (2021)
  283--295.

\bibitem{han2019joint}
M.~Han, J.~Kim, Joint banknote recognition and counterfeit detection using
  explainable artificial intelligence, Sensors 19~(16) (2019) 3607.

\bibitem{neupane2022temporal}
S.~Neupane, I.~A. Fernandez, W.~Patterson, S.~Mittal, S.~Rahimi, A temporal
  anomaly detection system for vehicles utilizing functional working groups and
  sensor channels, IEEE International Conference on Collaboration and Internet
  Computing (IEEE CIC 2022) (2022).

\bibitem{darpa2016broad}
DARPA, Broad agency announcement explainable artificial intelligence (xai),
  DARPA-BAA-16-53 (2016) 7--8.

\bibitem{islam2019domain}
S.~R. Islam, W.~Eberle, S.~K. Ghafoor, A.~Siraj, M.~Rogers, Domain knowledge
  aided explainable artificial intelligence for intrusion detection and
  response, arXiv preprint arXiv:1911.09853 (2019).

\bibitem{wu2020feature}
C.~Wu, A.~Qian, X.~Dong, Y.~Zhang, Feature-oriented design of visual analytics
  system for interpretable deep learning based intrusion detection, in: 2020
  International Symposium on Theoretical Aspects of Software Engineering
  (TASE), IEEE, 2020, pp. 73--80.

\bibitem{wang2020explainable}
M.~Wang, K.~Zheng, Y.~Yang, X.~Wang, An explainable machine learning framework
  for intrusion detection systems, IEEE Access 8 (2020) 73127--73141.

\bibitem{khan2021new}
I.~A. Khan, N.~Moustafa, D.~Pi, K.~M. Sallam, A.~Y. Zomaya, B.~Li, A new
  explainable deep learning framework for cyber threat discovery in industrial
  iot networks, IEEE Internet of Things Journal (2021).

\bibitem{amarasinghe2018toward}
K.~Amarasinghe, K.~Kenney, M.~Manic, Toward explainable deep neural network
  based anomaly detection, in: 2018 11th International Conference on Human
  System Interaction (HSI), IEEE, 2018, pp. 311--317.

\bibitem{rumelhart1985feature}
D.~E. Rumelhart, D.~Zipser, Feature discovery by competitive learning,
  Cognitive science 9~(1) (1985) 75--112.

\bibitem{sammut2011encyclopedia}
C.~Sammut, G.~I. Webb, Encyclopedia of machine learning, Springer Science \&
  Business Media, 2011.

\bibitem{lichodzijewski2002host}
P.~Lichodzijewski, A.~N. Zincir-Heywood, M.~I. Heywood, Host-based intrusion
  detection using self-organizing maps, in: Proceedings of the 2002
  International Joint Conference on Neural Networks. IJCNN'02 (Cat. No.
  02CH37290), Vol.~2, IEEE, 2002, pp. 1714--1719.

\bibitem{de2015implementation}
E.~De~la Hoz, A.~Ortiz~Garc{\'\i}a, J.~Ortega~Lopera, E.~M. De~La Hoz~Correa,
  F.~E. Mendoza~Palechor, Implementation of an intrusion detection system based
  on self organizing map (2015).

\bibitem{pachghare2009intrusion}
V.~Pachghare, P.~Kulkarni, D.~M. Nikam, Intrusion detection system using self
  organizing maps, in: 2009 International Conference on Intelligent Agent \&
  Multi-Agent Systems, IEEE, 2009, pp. 1--5.

\bibitem{rhodes2000multiple}
B.~C. Rhodes, J.~A. Mahaffey, J.~D. Cannady, Multiple self-organizing maps for
  intrusion detection, in: Proceedings of the 23rd national information systems
  security conference, MD Press Baltimore, 2000, pp. 16--19.

\bibitem{albayrak2005combining}
S.~Albayrak, C.~Scheel, D.~Milosevic, A.~Muller, Combining self-organizing map
  algorithms for robust and scalable intrusion detection, in: International
  Conference on Computational Intelligence for Modelling, Control and
  Automation and International Conference on Intelligent Agents, Web
  Technologies and Internet Commerce (CIMCA-IAWTIC'06), Vol.~2, IEEE, 2005, pp.
  123--130.

\bibitem{palomo2009self}
E.~J. Palomo, E.~Dom{\'\i}nguez, R.~M. Luque, J.~Munoz, A self-organized
  multiagent system for intrusion detection, in: International Workshop on
  Agents and Data Mining Interaction, Springer, 2009, pp. 84--94.

\bibitem{qu2019statistics}
X.~Qu, L.~Yang, K.~Guo, L.~Ma, T.~Feng, S.~Ren, M.~Sun, Statistics-enhanced
  direct batch growth self-organizing mapping for efficient dos attack
  detection, IEEE Access 7 (2019) 78434--78441.

\bibitem{ippoliti2012ghsom}
D.~Ippoliti, X.~Zhou, A-ghsom: An adaptive growing hierarchical self organizing
  map for network anomaly detection, Journal of Parallel and Distributed
  Computing 72~(12) (2012) 1576--1590.

\bibitem{palomo2008new}
E.~J. Palomo, E.~Dom{\'\i}nguez, R.~M. Luque, J.~Mu{\~n}oz, A new ghsom model
  applied to network security, in: International Conference on Artificial
  Neural Networks, Springer, 2008, pp. 680--689.

\bibitem{salem2013enhanced}
M.~Salem, U.~Buehler, An enhanced ghsom for ids, in: 2013 IEEE International
  Conference on Systems, Man, and Cybernetics, IEEE, 2013, pp. 1138--1143.

\bibitem{yang2010using}
Y.~Yang, D.~Jiang, M.~Xia, Using improved ghsom for intrusion detection,
  Journal of Information Assurance and Security 5 (2010) 232--239.

\bibitem{bahrololum2008anomaly}
M.~Bahrololum, M.~Khaleghi, Anomaly intrusion detection system using gaussian
  mixture model, in: 2008 Third International Conference on Convergence and
  Hybrid Information Technology, Vol.~1, IEEE, 2008, pp. 1162--1167.

\bibitem{hammad2022mmm}
M.~Hammad, N.~Hewahi, W.~Elmedany, Mmm-rf: A novel high accuracy multinomial
  mixture model for network intrusion detection systems, Computers \& Security
  (2022) 102777.

\bibitem{bitaab2017hybrid}
M.~Bitaab, S.~Hashemi, Hybrid intrusion detection: Combining decision tree and
  gaussian mixture model, in: 2017 14th International ISC (Iranian Society of
  Cryptology) Conference on Information Security and Cryptology (ISCISC), IEEE,
  2017, pp. 8--12.

\bibitem{muda2011intrusion}
Z.~Muda, W.~Yassin, M.~Sulaiman, N.~Udzir, Intrusion detection based on k-means
  clustering and na{\"\i}ve bayes classification, in: 2011 7th international
  conference on information technology in Asia, IEEE, 2011, pp. 1--6.

\bibitem{tahir2016oving}
H.~M. Tahir, A.~M. Said, N.~H. Osman, N.~H. Zakaria, P.~N.~M. Sabri, N.~Katuk,
  Oving k-means clustering using discretization technique in network intrusion
  detection system, in: 2016 3rd International Conference on Computer and
  Information Sciences (ICCOINS), IEEE, 2016, pp. 248--252.

\bibitem{li2011anomaly}
Z.~Li, Y.~Li, L.~Xu, Anomaly intrusion detection method based on k-means
  clustering algorithm with particle swarm optimization, in: 2011 international
  conference of information technology, computer engineering and management
  sciences, Vol.~2, IEEE, 2011, pp. 157--161.

\bibitem{oja1999kohonen}
E.~Oja, S.~Kaski, Kohonen maps, Elsevier, 1999.

\bibitem{guthikonda2005kohonen}
S.~M. Guthikonda, Kohonen self-organizing maps, Wittenberg University 98
  (2005).

\bibitem{kohonen2007kohonen}
T.~Kohonen, T.~Honkela, Kohonen network, Scholarpedia 2~(1) (2007) 1568.

\bibitem{yuan2018implementation}
L.~Yuan, Implementation of self-organizing maps with Python, University of
  Rhode Island, 2018.

\bibitem{Ong1999DataMU}
J.~Ong, S.~M.~R. Abidi, Data mining using self-organizing kohonen maps: A
  technique for effective data clustering \& visualization, in: IC-AI, 1999.

\bibitem{Breard2017}
G.~Breard, Evaluating self-organizing map quality measures as convergence
  criteria, 2017.

\bibitem{liu2018scalable}
Y.~Liu, J.~Sun, Q.~Yao, S.~Wang, K.~Zheng, Y.~Liu, A scalable heterogeneous
  parallel som based on mpi/cuda, in: Asian Conference on Machine Learning,
  PMLR, 2018, pp. 264--279.

\bibitem{Alahakoon1998}
D.~Alahakoon, S.~Halgamuge, B.~Srinivasan, A self-growing cluster development
  approach to data mining, in: SMC'98 Conference Proceedings. 1998 IEEE
  International Conference on Systems, Man, and Cybernetics (Cat.
  No.98CH36218), Vol.~3, 1998, pp. 2901--2906 vol.3.
\newblock \href {https://doi.org/10.1109/ICSMC.1998.725103}
  {\path{doi:10.1109/ICSMC.1998.725103}}.

\bibitem{vasighi2017directed}
M.~Vasighi, H.~Amini, A directed batch growing approach to enhance the topology
  preservation of self-organizing map, Applied Soft Computing 55 (2017)
  424--435.

\bibitem{Myles2004}
A.~J. Myles, R.~N. Feudale, Y.~Liu, N.~A. Woody, S.~D. Brown, An introduction
  to decision tree modeling, Journal of Chemometrics 18 (2004) 275--285.
\newblock \href {https://doi.org/10.1002/CEM.873} {\path{doi:10.1002/CEM.873}}.

\bibitem{Quinlan1987}
J.~R. Quinlan, Simplifying decision trees, International Journal of Man-Machine
  Studies 27 (1987) 221--234.
\newblock \href {https://doi.org/10.1016/S0020-7373(87)80053-6}
  {\path{doi:10.1016/S0020-7373(87)80053-6}}.

\bibitem{mansour1997pessimistic}
Y.~Mansour, Pessimistic decision tree pruning based on tree size, in: Machine
  Learning-International Workshop then Conference, Citeseer, 1997, pp.
  195--201.

\bibitem{Kearns1998AFB}
M.~Kearns, Y.~Mansour, A fast, bottom-up decision tree pruning algorithm with
  near-optimal generalization, in: ICML, 1998.

\bibitem{Hamel2012a}
L.~Hamel, C.~Brown, Bayesian probability approach to feature significance for
  infrared spectra of bacteria, Applied Spectroscopy 66 (2012) 48--59.
\newblock \href {https://doi.org/10.1366/10-06155}
  {\path{doi:10.1366/10-06155}}.

\bibitem{Tavallaee2009}
M.~Tavallaee, E.~Bagheri, W.~Lu, A.~A. Ghorbani, A detailed analysis of the kdd
  cup 99 data set, 2009, pp. 1--6.
\newblock \href {https://doi.org/10.1109/CISDA.2009.5356528}
  {\path{doi:10.1109/CISDA.2009.5356528}}.

\bibitem{sharafaldin2018toward}
I.~Sharafaldin, A.~H. Lashkari, A.~A. Ghorbani, Toward generating a new
  intrusion detection dataset and intrusion traffic characterization., ICISSp 1
  (2018) 108--116.

\bibitem{carvalho2014heartbleed}
M.~Carvalho, J.~DeMott, R.~Ford, D.~A. Wheeler, Heartbleed 101, IEEE security
  \& privacy 12~(4) (2014) 63--67.

\bibitem{lashkari2017cicflowmeter}
A.~H. Lashkari, Y.~Zang, G.~Owhuo, M.~Mamun, G.~Gil, Cicflowmeter (2017).

\bibitem{Kohonen1998}
T.~Kohonen, The self-organizing map, Neurocomputing 21 (1998) 1--6.
\newblock \href {https://doi.org/https://doi.org/10.1016/S0925-2312(98)00030-7}
  {\path{doi:https://doi.org/10.1016/S0925-2312(98)00030-7}}.

\bibitem{Lampinen1992}
J.~Lampinen, E.~Oja, \href{https://doi.org/10.1007/BF00118594}{Clustering
  properties of hierarchical self-organizing maps}, Journal of Mathematical
  Imaging and Vision 2 (1992) 261--272.
\newblock \href {https://doi.org/10.1007/BF00118594}
  {\path{doi:10.1007/BF00118594}}.
\newline\urlprefix\url{https://doi.org/10.1007/BF00118594}

\bibitem{Hamel2016}
L.~Hamel, Som quality measures: An efficient statistical approach, Vol. 428,
  Springer Verlag, 2016, pp. 49--59.
\newblock \href {https://doi.org/10.1007/978-3-319-28518-4}
  {\path{doi:10.1007/978-3-319-28518-4}}.

\bibitem{Tatoian2018}
\href{https://doi.org/10.4018/IJSSMET.2018040103}{Self-organizing map
  convergence}, Int. J. Serv. Sci. Manag. Eng. Technol. 9 (2018) 61--84.
\newblock \href {https://doi.org/10.4018/IJSSMET.2018040103}
  {\path{doi:10.4018/IJSSMET.2018040103}}.
\newline\urlprefix\url{https://doi.org/10.4018/IJSSMET.2018040103}

\bibitem{wickramasinghe2021explainable}
C.~S. Wickramasinghe, K.~Amarasinghe, D.~L. Marino, C.~Rieger, M.~Manic,
  Explainable unsupervised machine learning for cyber-physical systems, IEEE
  Access 9 (2021) 131824--131843.

\bibitem{Hamel2012}
L.~Hamel, C.~Brown, Improved interpretability of the unified distance matrix
  with connected components, 7th International Conference on Data Mining
  (DMIN'11) (4 2012).

\bibitem{jia2019network}
Y.~Jia, M.~Wang, Y.~Wang, Network intrusion detection algorithm based on deep
  neural network, IET Information Security 13~(1) (2019) 48--53.

\bibitem{mohammadpour2018convolutional}
L.~Mohammadpour, T.~C. Ling, C.~S. Liew, C.~Y. Chong, A convolutional neural
  network for network intrusion detection system, Proceedings of the
  Asia-Pacific Advanced Network 46~(0) (2018) 50--55.

\bibitem{xu2018intrusion}
C.~Xu, J.~Shen, X.~Du, F.~Zhang, An intrusion detection system using a deep
  neural network with gated recurrent units, IEEE Access 6 (2018) 48697--48707.

\bibitem{su2020bat}
T.~Su, H.~Sun, J.~Zhu, S.~Wang, Y.~Li, Bat: Deep learning methods on network
  intrusion detection using nsl-kdd dataset, IEEE Access 8 (2020) 29575--29585.

\bibitem{khan2021spectrogram}
A.~S. Khan, Z.~Ahmad, J.~Abdullah, F.~Ahmad, A spectrogram image-based network
  anomaly detection system using deep convolutional neural network, IEEE Access
  9 (2021) 87079--87093.

\bibitem{almutlaq2022two}
S.~Almutlaq, A.~Derhab, M.~M. Hassan, K.~Kaur, Two-stage intrusion detection
  system in intelligent transportation systems using rule extraction methods
  from deep neural networks, IEEE Transactions on Intelligent Transportation
  Systems (2022).

\bibitem{abdel2021semi}
M.~Abdel-Basset, H.~Hawash, R.~K. Chakrabortty, M.~J. Ryan, Semi-supervised
  spatiotemporal deep learning for intrusions detection in iot networks, IEEE
  Internet of Things Journal 8~(15) (2021) 12251--12265.

\bibitem{halbouni2022cnn}
A.~H. Halbouni, T.~S. Gunawan, M.~Halbouni, F.~A.~A. Assaig, M.~R. Effendi,
  N.~Ismail, Cnn-ids: Convolutional neural network for network intrusion
  detection system, in: 2022 8th International Conference on Wireless and
  Telematics (ICWT), IEEE, 2022, pp. 1--4.

\end{thebibliography}





% \end{thebibliography}
%%
%% End of file `elsarticle-template-num.tex'.

\end{document}